\documentclass{article}

\usepackage{arxiv}

\usepackage{enumitem}
\usepackage{color}
\usepackage{array}
\usepackage{graphicx}
\usepackage{bbding}
\usepackage{amssymb}
\usepackage{booktabs}
\usepackage{hyperref}
\usepackage{multirow}
\usepackage{bigstrut}
\usepackage{threeparttable}
\usepackage{listings}
\usepackage{diagbox}
\usepackage{float}

\newcommand{\tab}{\hspace*{1em}}
\newcommand{\code}[1]{{\fontfamily{cmtt}\fontseries{m}\fontshape{n}\selectfont\small{#1}}}
\newcommand{\system}{\code{EthScope}}
\newcommand{\txspector}{\textsc{TxSpector}}
\newcommand{\aggregator}{\code{data aggregator}}
\newcommand{\Aggregator}{\code{Data Aggregator}}

\newcommand{\engine}{\code{replay engine}}
\newcommand{\Engine}{\code{Replay Engine}}
\newcommand{\framework}{\code{instrumentation framework}}
\newcommand{\Framework}{\code{Instrumentation Framework}}

\newcommand{\minitab}[2][l]{\begin{tabular}{#1}#2\end{tabular}}

\usepackage[numbers,sort]{natbib}

\newcommand{\tabincell}[2]{\begin{tabular}{@{}#1@{}}#2\end{tabular}}

\usepackage{listings, xcolor}

\definecolor{verylightgray}{rgb}{.97,.97,.97}

\lstdefinelanguage{Solidity}{
	keywords=[1]{anonymous, assembly, assert, balance, break, call, callcode, case, catch, class, constant, continue, constructor, contract, debugger, default, delegatecall, delete, do, else, emit, event, experimental, export, external, false, finally, for, function, gas, if, implements, import, in, indexed, instanceof, interface, internal, is, length, library, log0, log1, log2, log3, log4, memory, modifier, new, payable, pragma, private, protected, public, pure, push, require, return, returns, revert, selfdestruct, send, solidity, storage, struct, suicide, super, switch, then, this, throw, transfer, true, try, typeof, using, value, view, while, with, addmod, ecrecover, keccak256, mulmod, ripemd160, sha256, sha3}, %
	keywordstyle=[1]\color{blue}\bfseries,
	keywords=[2]{address, bool, byte, bytes, bytes1, bytes2, bytes3, bytes4, bytes5, bytes6, bytes7, bytes8, bytes9, bytes10, bytes11, bytes12, bytes13, bytes14, bytes15, bytes16, bytes17, bytes18, bytes19, bytes20, bytes21, bytes22, bytes23, bytes24, bytes25, bytes26, bytes27, bytes28, bytes29, bytes30, bytes31, bytes32, enum, int, int8, int16, int24, int32, int40, int48, int56, int64, int72, int80, int88, int96, int104, int112, int120, int128, int136, int144, int152, int160, int168, int176, int184, int192, int200, int208, int216, int224, int232, int240, int248, int256, mapping, string, uint, uint8, uint16, uint24, uint32, uint40, uint48, uint56, uint64, uint72, uint80, uint88, uint96, uint104, uint112, uint120, uint128, uint136, uint144, uint152, uint160, uint168, uint176, uint184, uint192, uint200, uint208, uint216, uint224, uint232, uint240, uint248, uint256, var, void, ether, finney, szabo, wei, days, hours, minutes, seconds, weeks, years},	%
	keywordstyle=[2]\color{teal}\bfseries,
	keywords=[3]{block, blockhash, coinbase, difficulty, gaslimit, number, timestamp, msg, data, gas, sender, sig, value, now, tx, gasprice, origin},	%
	keywordstyle=[3]\color{violet}\bfseries,
	identifierstyle=\color{black},
	sensitive=false,
	comment=[l]{//},
	morecomment=[s]{/*}{*/},
	commentstyle=\color{gray}\ttfamily,
	stringstyle=\color{red}\ttfamily,
	morestring=[b]',
	morestring=[b]"
}

\usepackage{listings, xcolor}

\definecolor{verylightgray}{rgb}{.97,.97,.97}

\lstdefinelanguage{JavaScript}%
  {morekeywords={typeof,new,true,false,catch,function,return,null,catch,switch,var,if,in,while,do,else,case,break},%
  morecomment=[l]//,%
  morecomment=[s]{/*}{*/},%
  morestring=[b]",%
  morestring=[b]',%
  }[keywords,comments,strings]%
  
\usepackage[affil-it]{authblk}

\usepackage{blindtext}

\begin{document}
	%
	\title{Time-Travel Investigation: Towards Building A Scalable Attack Detection Framework on Ethereum}

	\author[1]{Lei Wu}
	\author[1]{Siwei Wu}
	\author[1]{Yajin Zhou\thanks{Corresponding author (yajin\_zhou@zju.edu.cn).}\hspace*{0.4em}}
	\author[1]{Runhuai Li}
	\author[2]{Zhi Wang}
	\author[3]{Xiapu Luo}
	\author[4]{Cong Wang}
	\author[1]{Kui Ren}
	\affil[1]{Zhejiang University}
	\affil[2]{Florida State University}
	\affil[3]{The Hong Kong Polytechnic University}
	\affil[4]{City University of Hong Kong}

	\maketitle

	\begin{abstract}
		
		As one of the representative blockchain platforms, Ethereum has attracted lots of attacks. Due to the existed financial loss, there is a pressing need to perform timely investigation and detect more attack instances. Though multiple systems have been proposed, they suffer from the scalability issue due to the following reasons. First, the tight coupling between malicious contract detection and blockchain data importing makes them infeasible to repeatedly detect different attacks. Second, the coarse-grained archive data makes them inefficient to replay transactions. Third, the separation between malicious contract detection and runtime state recovery consumes lots of storage. 
		
		In this paper, we present the design of a scalable attack detection framework on Ethereum. It overcomes the scalability issue by saving the Ethereum state into a database and providing an efficient way to locate suspicious transactions. The saved state is fine-grained to support the replay of arbitrary transactions. The state is well-designed to avoid saving unnecessary state to optimize the storage consumption. We implement a prototype named \system{} and solve three technical challenges, i.e., incomplete Ethereum state, scalability, and extensibility. The performance evaluation shows that our system can solve the scalability issue, i.e., efficiently performing a large-scale analysis on billions of transactions, and a speedup of around 2,300x when replaying transactions. It also has lower storage consumption compared with existing systems. The result with three different types of information as inputs shows that our system can help an analyst understand attack behaviors and further detect more attacks. To engage the community, we will release our system and the dataset of detected attacks.
		
	\end{abstract}
	
	\section{Introduction}
	\label{sec:introduction}
	
	With an explosive growth of the blockchain technique, Ethereum~\cite{ethereum} has become one 
	of the representative platforms. One reason is due to its inborn support of smart contracts.
	Developers use smart contracts to build Decentralized Applications (DApps), ranging from gaming,
	lottery, Decentralized Finance (DeFi), and cryptocurrency, e.g., ERC20 tokens~\cite{erc20tokenstandard}.
	
	At the same time, attacks targeting Ethereum are increasing.
	By exploiting the vulnerabilities of smart contracts,
	attackers could make huge profits in a short time. For instance, in April 2016,
	attackers exploited the 
	re-entrancy vulnerability in the DAO smart contract and stole around 3.6 million Ether~\cite{dao}.
	Attackers used the similar vulnerability to attack the decentralized exchange Uniswap~\cite{Uniswap} (July 2019)
	and DeFi application Lend.Me~\cite{Lend.Me} (April 2020). Besides, lots of other types of attacks have been observed in the wild~\cite{batchoverflow, proxyoverflow, ceoanyone}.
	
	\begin{figure}[t]
		\centering
		\includegraphics[width=0.45\textwidth]{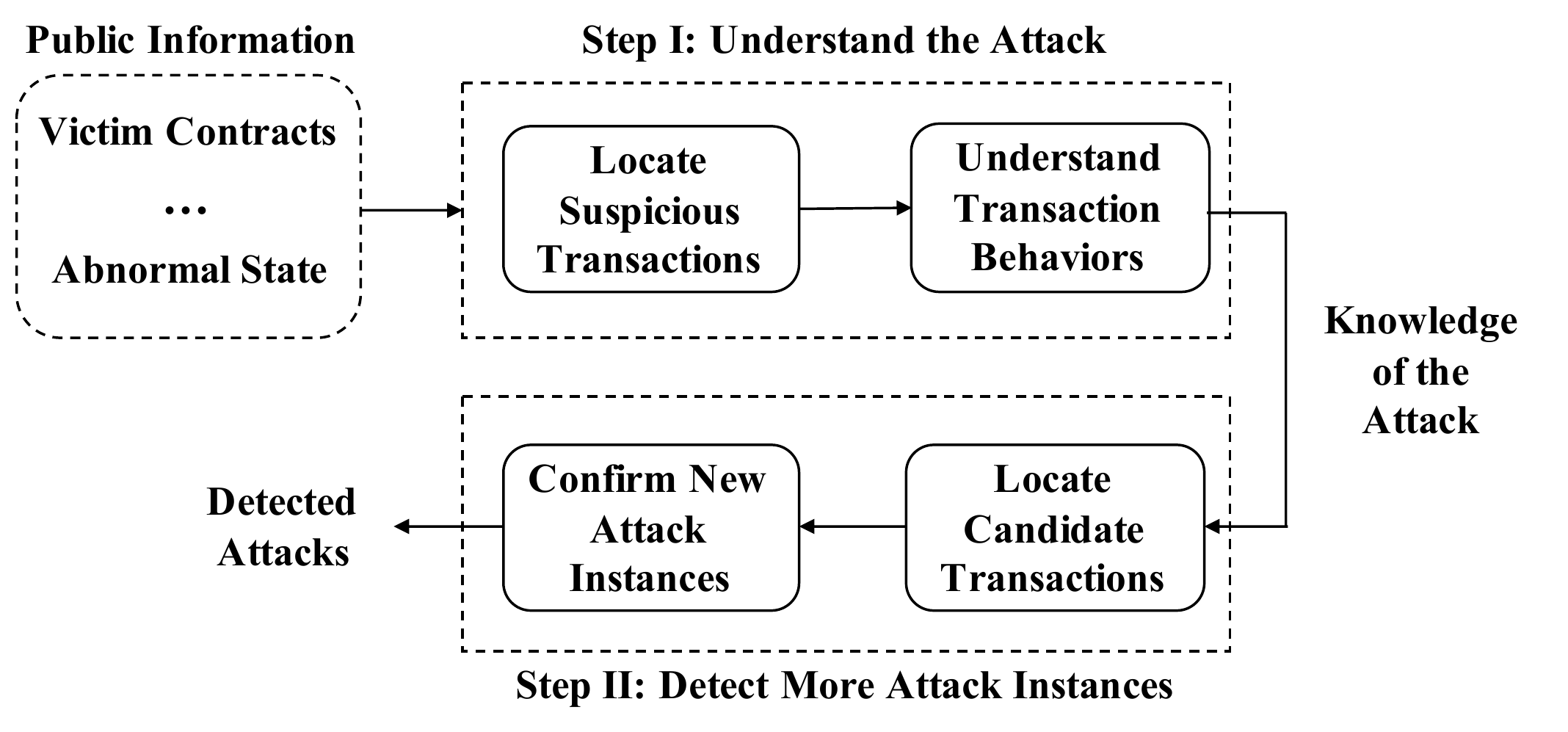}
		\caption{The typical flow of an investigation of attacks on Ethereum.}
		\label{fig:investigation}
		\vspace{-0.2in}
	\end{figure}

	Accordingly, there is a pressing need for the security community to perform timely investigations on attacks and detect more attack instances that were not revealed. This requires the capability to quickly locate suspicious transactions based on various types of public information. For instance, suppose there is a reported attack to a smart contract (the victim contract) on the public forum, but the detail of such an attack is unknown. In order to understand the attack, an analyst needs to locate suspicious transactions that interact with the victim contract, and further construct the callgraph between the victim contract and others to understand their behaviors. After that, the analyst may need to detect more attack instances. In particular, he or she further locates candidate transactions~\footnote{To avoid the confusion with suspicious transactions used in step I, we call transactions that are potentially related to the attack in this step as candidate transactions.} that are potentially related to the attack and \textit{replay them}. By doing so, the analyst can monitor the runtime state of a smart contract and hook into its execution to detect more attacks. Fig.~\ref{fig:investigation} shows this flow.

	Note that, the investigation may continuously repeat the steps in Fig.~\ref{fig:investigation}. That's because the understanding of an attack needs multiple rounds of querying and analyzing transactions. This raises the challenge that \textit{the analysis framework should be scalable to a large number of transactions} (until July 5th, 2020, Ethereum has $754,614,255$ normal transactions and $962,171,044$ internal transactions, respectively), i.e., \textit{efficiently locating and replaying transactions.}~\footnote{Because the investigation involves the {replay} of transactions to monitor the Ethereum state, it is like a time travel to certain points in time, hence the name \textit{time-travel investigation}.}

	\smallskip 
	\noindent \textbf{Limitations of existing systems}\tab 
	Though multiple systems~\cite{Sereum, ECFChecker, ferreira2019aegis, chensoda, TXSPECTOR}  have been proposed to detect malicious smart contracts~\footnote{In this paper, we interchangeably use the following two terms, i.e., malicious smart contracts and attacks, because attacks are usually automatically performed by malicious smart contracts.}, the scalability issue makes them ineffective to perform the time-travel investigation due to the following reasons.
	
	\begin{itemize} [leftmargin=*]
		\item {\textit{Tight Coupling between malicious contract detection and blockchain data importing} \textbf{(Limitation I)}}\tab
		Some systems import the entire blockchain data from the genesis block and replay historical transactions. During this process, malicious contracts are detected based on pre-defined rules. The importing process is time-consuming (about ten days) and cannot be repeated. It is inflexible to repeatedly replay transactions, revise and debug detection rules, a considerable limitation to detect new attack instances.
		
		\smallskip
		\item {\textit{Coarse-grained archive data} \textbf{(Limitation II)}}\tab
		To solve the previous limitation, systems could leverage the \textit{archive mode}~\cite{archivedata} of popular Ethereum clients to repeatedly replay arbitrary transactions, after importing the data once. 
		However, the historical state is too coarse-grained to \textit{efficiently} replay transactions, since unnecessary transactions are executed (Section~\ref{subsec:app_challenges}). Our evaluation shows that it costs more than $47$ minutes to replay $100$ normal transactions. This is not scalable for real attack detection, which needs to replay tens of thousands and even millions of transactions (Section~\ref{subsubsec:eva_engine}).
		
		\smallskip
		\item {\textit{Separation between malicious contract detection and runtime state recovery} \textbf{(Limitation III)}}\tab 
		Instead of using the coarse-grained archive data, recent systems recover and store the runtime information (called logical relation in the paper~\cite{TXSPECTOR}) into a database. The further detection is based on the stored logical relation. This avoids the cost of repeatedly replaying transactions. However, the storage for the logical relation is huge. 
		For instance, the logical relation database for blocks ranging from $7,000,000$ to $7,200,000$ consumes $2,949$ GB~\cite{TXSPECTOR}. Given the fact that Ethereum has around $10,400,000$ blocks (as on July 5th, 2020) and this number is still increasing, it's not practical to detect attacks in the whole Ethereum blocks.

	\end{itemize}

	\smallskip 
	\noindent \textbf{Our approach}\tab 
	Our system takes the following approaches to overcome the limitations. 
	
	\begin{itemize} [leftmargin=*]
		\item
		{\textbf{Limitation I}}: Our system does not perform the detection during the blockchain importing process.
		Instead, we save the Ethereum state, e.g., internal transactions, created smart contract code, into a database. Further detection is based on the saved state to locate suspicious transactions. This decouples the detection and the importing process.
		
		\smallskip
		\item
		{\textbf{Limitation II}}: Our system replays arbitrary transactions in a scalable way. This is due to the well-designed and fine-grained state that has been retrieved in the previous step. By doing so, there is no need to reply unnecessary transactions in our system. For instance, our system only needs around one second to replay the same $100$ normal transactions that consumed $47$ minutes in the archive mode (Table~\ref{tab:eva_inst_performance}).
		
		\smallskip
		\item 
		{\textbf{Limitation III}}: Our detection is performed at the same time when replaying transactions. It provides an extensive way for an analyst to specify detection rules, which are executed when replaying transactions. Thanks to the efficient replay engine, our system does not need to save unnecessary runtime information. For instance, our system only consumes $1,844$ GB storage for the historical state in the \textit{$10.5$ million blocks} (as on July 22th, 2020), compared with $2,949$ GB needed for \textit{$0.2$ million blocks} in \txspector{}~\cite{TXSPECTOR}. This makes the detection among all Ethereum blocks possible.
	\end{itemize}

	\smallskip 
	\noindent \textbf{System Implementation}\tab 
	With the scalability requirement in mind, we have implemented an analysis framework named \system{} with three components.

	Specifically, the first component, i.e., \aggregator{}, collects and recovers the critical blockchain state, including internal transactions, self-destructed smart contracts, the account balance of each
	block, and etc. The database is used to quickly locate suspicious transactions, and more importantly, provide fine-grained state that is needed by the \engine{}.  
	
	The second component, i.e., \engine{}, is able to \textit{efficiently and repeatedly
		{replay} arbitrary and a large number of transactions}. This is critical  to solve the scalability issue in existing systems. The saved blockchain state is carefully designed to replay transactions without executing unnecessary ones.

	The third component, i.e., \framework{}, exposes interfaces for
	an analyst to dynamically instrument smart contracts
	and introspect the execution of transactions. An analyst can develop
	analysis scripts (using the JavaScript language) to analyze
	transactions and detect malicious smart contracts. 
	Our framework reduces the performance
	overhead by a fine-grained design of instrumentation points and minimizes
	context switches between the EVM and the analysis script.
	Compared with JSTracer~\cite{js-tracer}, our framework
	is more flexible and efficient (Table~\ref{tab:eva_inst_performance}).

	\smallskip 
	\noindent \textbf{Evaluation}\tab 
	We evaluate our system from two perspectives. 
	We first evaluate the efficiency of our system. The performance evaluation shows that our system solves the scalability problem. Specifically, our system consumes $1,817$ GB for the state of $10,400,000$ blocks. It is more efficient (around $2,300$x speedup) than existing ones when replaying transactions. 
	Then we use three different types of public information to detect attacks on Ethereum. Specifically, we leverage a victim smart contract, a reported suspicious transaction, and the abnormal blockchain state as inputs to understand the attack and further detect more attack instances. The comparison with our system and other ones on the detection of the re-entrancy attack shows the accuracy of our system.
	
	In summary, this paper makes the following main contributions:
	\begin{itemize} [leftmargin=*]
		\item We present the flow of an investigation of attacks on Ethereum and summarize the limitations of existing systems and their reasons.
		
		\item We propose multiple methods to solve the scalability issue and present the design of a scalable framework to detect \textit{real} attacks on Ethereum (Section~\ref{sec:approach}).
		
		\item We implement a prototype and illustrate methods to address three technical challenges  (Section~\ref{sec:implementation}).
		
		\item We evaluate the performance and effectiveness of our system with comprehensive experiments (Section~\ref{sec:evaluation}).
	\end{itemize}

	To engage the community, we will release the source code of \system{}. 
	We have released a trial system with a Docker image on \url{https://hub.docker.com/r/swaywu/ethscope-trial}.
	
	\section{Background}
	\label{sec:background}
	
	\subsection{Ethereum Accounts}
	\label{subsec:b_accounts}
	Each account in Ethereum has an address and associated balance
	in {Ether}. There exist two types of accounts, i.e.,
	externally owned account (EOA) and smart contract account, respectively.
	EOAs are controlled by private keys, while smart contract accounts
	are controlled by their contract code~\cite{wp-ethereum}.
	Note that, both accounts can have Ether and other tokens, thus
	are associated with balances~\footnote{A smart contract
		account can have balances may contradict one's intuition.}.
	
	The address of a new smart contract is calculated from 
	the number of transactions being sent (nonce) and  the address of
	its creator, which is the account that creates the smart contract.
	Due to this, the newly created contract address
	is predictable by its creator.
	We will illustrate an attack that exploits this property in
	Section~\ref{subsec:eva_victim_contract}.
	
	\begin{figure}[t]
	\centering
    \includegraphics[width=0.48\textwidth]{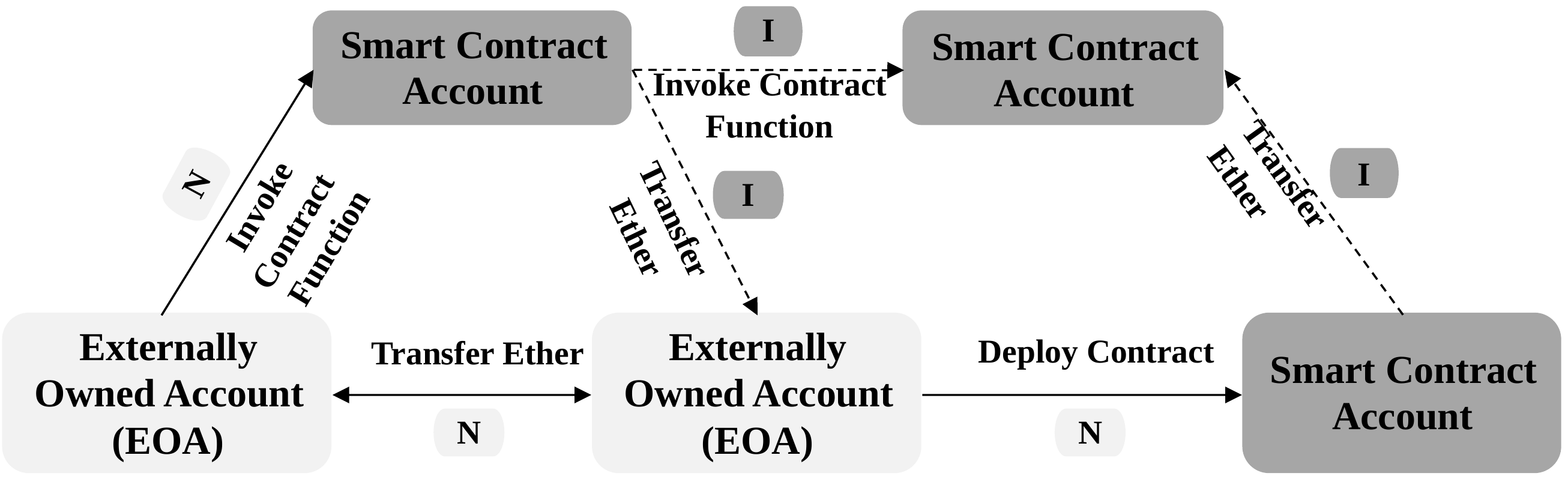}
    \caption{Normal and internal transactions. N: normal
    transactions; I: internal transactions.}
    \label{fig:b_transactions}
    \vspace{-0.2in}
\end{figure}

	\subsection{Transactions}
	\label{subsec:b_transactions}
	A transaction is a type of message call that serves three
	purposes, including transferring Ether, deploying a smart contract, and invoking
	functions of a smart contract. Transactions on the Ethereum are normally 
	initiated from EOAs, hence the name \textit{normal transactions}.
	
	Besides, there exists another type of transactions that are initiated from
	a smart contract. They are called \textit{internal transactions},
	which are used to invoke functions inside another smart contract, or transfer
	Ether to other accounts. 
	For instance, the opcode \code{CALL}
	can be used to invoke a function of another smart contract, thus creating an
	internal transaction. 
	
	Note that, an internal transaction is always
	initiated from a normal transaction, since the smart contract that
	creates an internal transaction should be executed in the first place
	(from an EOA using a normal transaction.) Moreover, a normal transaction
	could create numerous internal transactions, if the invoked smart contract
	does so (invoking functions of other smart contracts.)
	Fig.~\ref{fig:b_transactions} shows an overview of normal and internal transactions. 
	
	\subsection{Ethereum State}
	\label{subsec:b_ethereumstates}
	Ethereum's nodes are devices participating in validating transactions. 
	There are four types of state in Ethereum, which are useful to analyze and replay transactions.
	They include block information, normal transaction information, internal transaction information and accounts, as shown in the following.
	
	\begin{enumerate} [leftmargin=*]
		\item \textit{Block information.} The block information includes block number, block hash, and etc.
		\item \textit{Normal transaction information.} The normal transaction information includes  addresses of the sender and the receiver,
		transaction hash, transaction data, transaction values, and etc.
		\item \textit{Internal transaction information.} The internal transaction information is basically the same as
		the normal transaction, plus the depth of the call stack of EVM.
		\item \textit{Account state.} The account state includes balance, nonce, code and storage of each account	(including EOAs and smart contract accounts).
	\end{enumerate}

	Normally, a full Ethereum node only permanently stores the block information, normal transaction information and the
	account state of the \textit{latest blocks}. When synchronizing from the network, users can specify an option, e.g,
	\code{-gcmode=archive} in Geth, to retain a snapshot of accounts' state for each block. With the time-serial 
	accounts' state, users can use the API \code{debug.trace\_transaction} to replay arbitrary transactions
	in the exact manner as it was executed on the network. However, this method is not scalable. We will discuss the
	way used in our system to improve the performance of the replay process in Section~\ref{subsec:app_challenges}.

	\subsection{Smart Contracts}
	\noindent\textbf{Ethereum virtual machine}\tab
	A smart contract is a program that runs on an underlying Ethereum virtual machine
	(EVM) to transit the global state of the Ethereum network.
	A smart contract is usually programmed using a high-level language, e.g., Solidity, and then is compiled into low-level machine
	instructions (called \textit{opcodes}), which will be fetched, decoded and
	executed by EVM.
	
	EVM is a stack-based virtual machine. It has a virtual stack with $1,024$
	elements. All computations are performed on
	the stack. It means the operands, the result of intermediate operations
	are stored on the stack. For instance, when executing the \code{ADD} opcode to
	add two operands, EVM will pop two values from the stack,
	add them together and then push the result on the stack. 
	
	Besides the stack, there are four other types of data locations in EVM,
	\textit{memory}, \textit{storage}, \textit{input field}, and \textit{ret field}.
	The memory, input data and ret field are used to
	store temporary data such as function arguments, local variables, and
	return values. 
	They are volatile, which means their values will be
	lost when the execution of a smart contract is finished.
	In contrast, the storage is a (per-account) persistent
	key-value store.
	For instance, a gaming smart
	contract could leverage the storage to maintain the balance of each player.

	\smallskip \noindent\textbf{Function invocation}\tab
	As discussed in Section~\ref{subsec:b_transactions}, internal transactions
	are used to invoke smart contract functions.
	This is achieved through executing a message call~\cite{yp-ethereum}
	launched by six opcodes, including \code{CALL}, \code{CALLCODE},
	\code{DELEGATECALL}, \code{STATICCALL}, \code{CREATE} and \code{CREATE2}.
	
	In a smart contract, there is a signature (four bytes) to denote
	the destination function that will be invoked. 
	The signature is defined as the first four bytes
	of the hash value (SHA3) of the canonical representation of the function,
	including the function name and the parenthesized list of parameter
	types. Since this is a one-way function, it is hard to retrieve the function name
	from the signature. However, there is an online service~\cite{4byte}
	that we can lookup the function name given a signature.
	
	\smallskip \noindent\textbf{Smart contract creation and destruction}\tab
	A smart contract could be created using two opcodes, i.e., \code{CREATE}
	and \code{CREATE2}. Both opcodes behave similarly, except the way to calculate
	the address of the newly created smart contract~\cite{eip-1014}.
	
	A smart contract can be self-destructed through the opcode \code{SELFDESTRUCT}.
	This opcode destroys the smart contract itself, and transfers all the Ether
	inside the contract to the address specified in the parameters of this
	opcode (the target address). However, if the account with the target address
	does not exist, this opcode will create a new account with this address.
	This means that the \code{SELFDESTRUCT} opcode implicitly creates a new account. Moreover, self-destruction a smart contract reclaims the gas since it frees the resources on the blockchain.

	\section{System Design}
	\label{sec:approach}
	In the following, we will first illustrate technical challenges and then present the overall design of \system{}.
	
	\subsection{Technical Challenges}
	\label{subsec:app_challenges}
	
	There are three technical challenges for building a scalable attack detection framework on Ethereum.

	\begin{table}[t]
	\centering
	\caption{The comparison of state that could be retrieved by existing systems. Block: block information; NT: Normal transaction information; IT: internal transaction information; Account: Account state (Section~\ref{subsec:b_ethereumstates}). $\checkmark$: support; $\times$: not support; $\triangle$: partial support.}
	\label{tab:imp_hookpoints}
	\small
	\resizebox{0.5\columnwidth}{!}{%
	\begin{tabular}{|c|c|c|c|c|c|}
		\hline
		& \tabincell{c}{Block} & \tabincell{c}{NT} & \tabincell{c}{IT} & \tabincell{c}{Account} & \tabincell{c}{Interface} \\
		\hline
		  Ethereum full node   & $\checkmark$ & $\checkmark$ & $\times$ & $\times$ & $\times$ \\
	 	Archive node~\cite{archivedata} & $\checkmark$ & $\checkmark$ & $\times$ & $\triangle^{\rm a}$ & $\times$ \\
		Etherscan~\cite{Etherscan} & $\checkmark$ & $\checkmark$ & $\triangle^{\rm b}$ & $\triangle^{\rm c}$ &  $\triangle^{\rm d}$ \\
		BigQuery~\cite{bigqueryetl} & $\checkmark$ & $\checkmark$ & $\checkmark$  & $\times$ & $\checkmark$ \\
		Our system  & $\checkmark$ & $\checkmark$ & $\checkmark$  & $\checkmark$ & $\checkmark$ \\
		\hline
	\end{tabular}
}
 \begin{tablenotes}
 	\item $\triangle^{\rm a}$: The account state is coarse-grained that unnecessary transactions will be replayed (Section~\ref{subsubsec:eva_engine}).
 	\item $\triangle^{\rm b}$: Etherscan does not provide the invocation data of internal transactions.
 	\item $\triangle^{\rm c}$: The account state provided by Etherscan does not support the replay of transactions.
 	\item $\triangle^{\rm d}$: Etherscan does not support customized query for a large number of transactions, such as SQL.
 \end{tablenotes}

\label{tab:apch_datacomparsion}
\vspace{-0.2in}
\end{table}

	\smallskip \noindent \textbf{Incomplete blockchain state}\tab
	First, our system needs to provide a flexible interface to query the Ethereum state. For instance, when being used to understand and detect an attack, our system shall have the capability to quickly locate suspicious transactions from different perspectives, e.g., the values in the transactions or the number of internal transactions that exceed a certain threshold.
	Although there exist many methods that could be leveraged
	to explore Ethereum state, few of them fulfill our requirements. The details are shown in
	TABLE~\ref{tab:apch_datacomparsion}. 
	Among them, Ethereum in BigQuery~\cite{bigqueryetl} maintains the
	Ethereum state into seven tables and provides an SQL interface to query the state. However, it lacks the account state that is critical for replaying transactions.

	\smallskip \noindent \textbf{Scalability}\tab
	Our system needs to replay and analyze a large number of transactions. There exist three different methods that are adopted by existing systems~\cite{ECFChecker, ferreira2019aegis, chensoda}. All of them suffer from the
	scalability issue.
	
	The first one is to import the whole blockchain data with a customized EVM,
	which will execute all transactions (normal and internal ones) from the 
	genesis block (the first block on the chain). During this process,
	attack-specific rules are executed.
	Representative tools include ECFChecker~\cite{ECFChecker}, ÆGIS~\cite{ferreira2019aegis} and SODA~\cite{chensoda}.
	This method cannot selectively replay interested transactions.
	Thus, many unrelated ones have been
	executed, consuming lots of time. Moreover, the coupling between the detection and
	the  importing process makes the detection of new attack instances hard, since the time-consuming
	importing process cannot be executed repeatedly.

	The second way is to use the \code{debug.trace\_transaction} API~\cite{traceTransaction}
	exposed by Geth~\cite{go-ethereum} to \textit{replay} a  {transaction}
	with the Ethereum archive node~\cite{Sereum}. 
	Though this method is more efficient
	than the previous one, it still suffers from the
	scalability issue. That's because the granularity of historical state
	maintained by the Ethereum archive node is a block  
	rather than a transaction. In order to replay a
	transaction, all the (unnecessary) transactions before it inside the
	same block will be executed. 
	Our system solves this challenge by recovering a transaction-level historical state.

	The third one is first replaying all transactions and recording all the runtime information~\cite{TXSPECTOR}. The following detection is on the recorded information. However, this method consumes lots of storage.  According to the data reported in the paper~\cite{TXSPECTOR}, performing the attack detection in $2$ millions blocks cost at least $2,949$ GB. It's not scalable to analyze all the Ethereum blocks (more than $10$ millions blocks).

	\smallskip \noindent \textbf{Extensibility to detect different attacks}\tab
	Our system should be extensible to detect various attacks with analyst-provided scripts. 
	Geth has a mechanism called JSTracer~\cite{js-tracer} to introspect
	the execution of a smart contract.
	It allows users to specify a JavaScript file that will be invoked
	for every opcode executed.
	However, frequent switches between the EVM and the JavaScript file
	make it impractical to analyze a large number of transactions.
	Our system addresses this challenge with two optimizations.
	First, it has well-defined instrumentation points to minimize the number of
	context switches.
	The analysis script will be invoked on-demand (instead of each opcode) when
	defined instrumentation points are hit.
	Second, our framework is equipped with a dynamic taint analysis engine \textit{inside the EVM}. Analysts do not need to implement their own taint engine usings
	JavaScript files, which further reduces the number of context switches.

	\begin{figure}[t]
    \centering
    \includegraphics[width=0.5\textwidth]{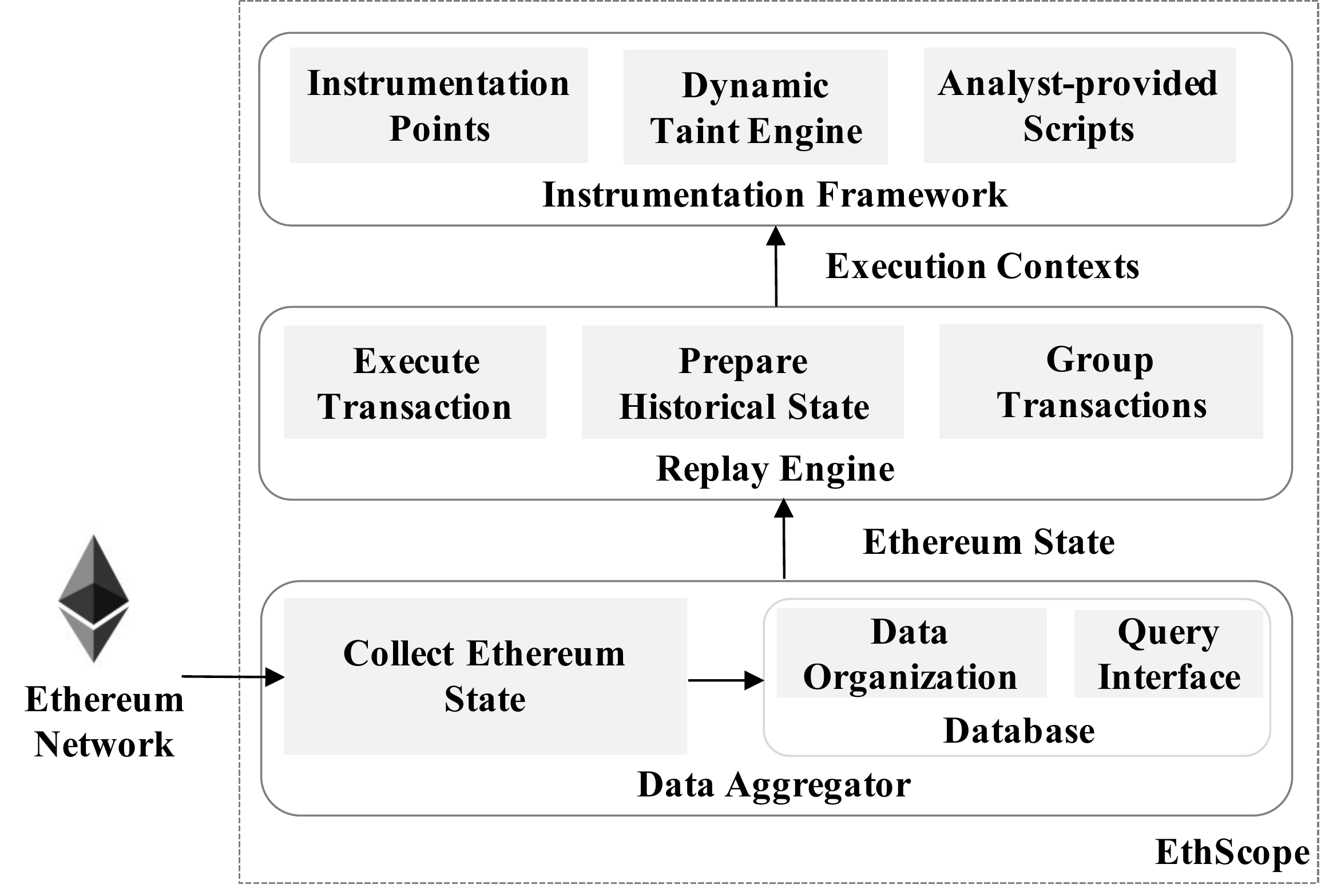}
    \caption{The overall architecture of  {\system}.}
    \label{fig:apch_architecture}
\vspace{-0.2in}
\end{figure}

	\subsection{Overall Design}
	We address these challenges with three components, i.e., \aggregator{}, \engine{}, and \framework{}. The overall system architecture is shown in Fig.~\ref{fig:apch_architecture}. 
	
	Specifically, \aggregator{} imports the whole blockchain data and collects the Ethereum state. The Ethereum state is collected by modifying the EVM. The collected state is stored in a cluster database equipped with a flexible query interface. An analyst could perform customized queries to locate transactions that are needed for further analysis. 
	Note that, the process to import the blockchain data is a one-time effort. All the saved state could be queried without the need to import the blockchain data again. Our system also takes a careful design of the stored state to save the storage consuming. In fact, it consumes less storage than the Ethereum archive mode (Section~\ref{subsubsec:eva_dataaggregator}).

	The second component, i.e, \engine{}, is used to replay arbitrary transactions. An analyst first locates candidate transactions and then feeds them to the engine. The \engine{} obtains the related state including related accounts' state for each transaction from the \aggregator{}. After that, it re-executes
	the transactions. Thanks for the transaction-level Ethereum state recovered by the \aggregator{}, our system does not need replay unnecessary transactions (Section~\ref{subsubsec:eva_engine}).
	
	The third component, i.e, \framework{}, provides a mechanism to customize the analysis. Specifically,  an analyst can develop analysis scripts
	by defining callback functions for instrumentation points.
	For instance, a specific callback function could be defined and will be invoked
	if and only if the \code{CALL} opcode is executed. 
	By doing so, our system avoids unnecessary context switches between EVM
	and the analysis script.
	During this process, the EVM state, including related stack
	and memory values, is provided to
	the script. Moreover, to facilitate the analysis, a dynamic taint engine is provided with well-defined APIs.

	\section{Implementation Details}
	\label{sec:implementation}
	
	We have implemented a prototype named \system{}. The \aggregator{} is implemented
	with around $1,137$ lines changes to the Geth client. Our system uses the distributed search and
	analytics engine Elasticsearch~\cite{elasticsearch} to store the Ethereum state and provide an
	interface to query them. The \engine{} and \framework{} are implemented with $5,191$ lines 
	changes to EVM. In the following, we will elaborate the implementation of each component.
	
	\subsection{\Aggregator{}}
	\label{subsec:dataaggregator}
	
	\smallskip \noindent \textbf{State collection}\tab
	The collection of block information and normal transaction information is straightforward.
	Our system changes the EVM to collect the data before the execution of each block (block information) and after the execution of each normal transaction (normal transaction information).
	
	Collecting internal transaction information and accounts' state requires our system to hook into
	the process of executing smart contracts. 
	For instance, when the opcode \code{SSTORE} is executed,
	the method \code{setState} in EVM is triggered.
	We change this method and add the code to capture the state.
	Note that, the state is not immediately stored into the underlying
	database. Instead,
	we create a buffer and save the state into the database when the buffer
	is full.
	
	One challenge is how to ensure the completeness and correctness of the collected state.
	In our system, we solve this challenge by comparing the collected state
	with ground truths. Specifically, for block information and normal transaction information, we can easily compare them with the data stored inside the Ethereum full node.
	For internal transaction information, we compare our data with the data provided by online services, e.g., Etherscan~\cite{Etherscan}.
	However, there are no ground truths for the transaction-grained historical accounts' state.
	We solve it in the \engine{} (State verification in Section~\ref{subsec:engine}).
	
	\smallskip \noindent \textbf{Data organization and query interface}\tab
	Our system takes the following methods to avoid the scalability issue caused by storage-consuming, while providing enough information to replay a transaction.
	Global variables of smart contracts consume lots of storage. That's because they are updated frequently in different blocks. 
	
	Theoretically, we need to store all the global variables for each transaction in each block. However, when replaying
	a transaction, only the variables touched by that transaction are needed. Thus, for each transaction, we only store the used global variables (storage values in Ethereum) in the database.

	Table~\ref{tab:app_elasticindices} in Appendix shows the detailed data schema. Specifically, the \code{Code} 
	index~\footnote{The index in Elasticsearch is similar to the database in a relational database.} stores
	the smart contracts' code and the \code{State} index records the information about creating and destructing accounts.
	Remaining ones are stored in the \code{Block} index. 
	
	Thanks to the Elasticsearch, an analyst could leverage the Query DSL
	based on JSON to define queries~\cite{elasticdsl} for customized analysis.
	
	\subsection{\Engine{}}
	\label{subsec:engine}
	In order to monitor the transaction behaviors at the runtime, we build an engine that is capable of replaying arbitrary transactions on Ethereum.
	Our engine is based on EVM of Geth, with modifications to add support to retrieve the state from \aggregator{}.
	Moreover, it provides interfaces to communicate with \framework{} (Section~\ref{subsec:framework}).
	
	\smallskip \noindent \textbf{Group transactions}\tab
	The input to \engine{} is a list of hash values for the transactions to be analyzed. In order to speed up the process of obtaining
	related data from \aggregator{}, our system divides transactions into different groups, with a threshold that each group contains no more than $10,000$ transactions. This threshold is related to the size of the system memory. 
	For each group, \engine{} first retrieves the historical state in a batch, and then replays transactions in the group.
	
	\smallskip \noindent \textbf{Retrieve Ethereum historical state}\tab
	In order to replay a normal transaction, we need to retrieve the Ethereum historical state from \aggregator{}. First, we get the block and 
	transaction information such as \code{Difficulty} and \code{GasLimit} from the \code{Block} index. Second, we retrieve the code
	of smart contracts that are related to this normal transaction in the nested field \code{GetCodeList} inside the field 
	\code{Transactions}. That's because a normal transaction could involve \textit{multiple smart contracts}. We retrieve the code for all the
	smart contracts. Third, we obtain all accounts' state: nonce, balance and storage values that the transaction will load. 
	When the normal transaction is to create a new smart contract, we need to retrieve the deploying code of the new smart contract
	from the index \code{Code}, which is the input of this normal transaction, too. Table~\ref{tab:app_elasticindices} in Appendix
	shows the details of the mentioned fields and indices.
	
	\smallskip \noindent \textbf{Execute transactions}\tab
	After retrieving the historical state, \engine{} executes the transactions. During this process, callback functions defined in the
	analysis script will be invoked.
	In order to speed up the process, our system further divides transactions in a group
	into different clusters according to the number of CPU cores, and executes transactions inside different clusters in parallel.
	
	\smallskip \noindent \textbf{Verify state}\tab
	After replaying each normal transaction, \engine{} will compare the used gas and output of this transaction with the same fields in
	the normal transaction information in \aggregator{}. This ensures the correctness of the replay process. Note that, the normal transaction information in \aggregator{} has been verified (State collection in Section~\ref{subsec:dataaggregator}). 
	
	\subsection{\Framework{}}
	\label{subsec:framework}
	The \framework{} aims to provide extensible APIs for an analyst to develop analysis scripts to detect new attack instances. Besides, \framework{}
	provides a dynamic taint engine to facilitate the analysis of control dependency and data dependency.
	
	\smallskip \noindent \textbf{Overview}\tab
	The framework is hooked into the \engine{} and provides JavaScript interfaces. Our system uses the Duktape JavaScript engine binding
	for Go~\cite{goduktape} to execute JavaScript functions inside the EVM developed in Go. Specifically, it defines
	\textit{instrumentation points}, where the replay process will be suspended and user-specific callback functions (in JavaScript) will be invoked. At
	the same time, it provides the interfaces for analysis scripts to access the current execution context, such as stack values and memory values. When the callback function finishes its execution, the \engine{} continues the smart contract's execution from the instruction after the instrumentation point.

	\begin{table}[t]
    \centering
    \caption{Three types of instrumentation points supported in our system.
    O: opcode-orientated;
    T: transaction-orientated; C: context-orientated.}
    \label{tab:imp_hookpoints}
    \small
    \resizebox{0.5\columnwidth}{!}{%
    \begin{tabular}{|c|c|c|}
        \hline
Instrumentation Points & Type & Description \\ \hline
\multicolumn{1}{|c|}{\begin{tabular}[c]{@{}c@{}}\{op\}\\ after\{Op\}\end{tabular}} & \multicolumn{1}{c|}{O} & \multicolumn{1}{c|}{\begin{tabular}[c]{@{}c@{}}before and after the opcode \{op\}\\ is executed\end{tabular}} \\ \hline
\multicolumn{1}{|c|}{\begin{tabular}[c]{@{}c@{}}transactionStart\\ transactionEnd\end{tabular}} & \multicolumn{1}{c|}{T} & \multicolumn{1}{c|}{\begin{tabular}[c]{@{}c@{}}before and after an external\\ transaction is executed\end{tabular}} \\ \hline
\multicolumn{1}{|c|}{\begin{tabular}[c]{@{}c@{}}contractStart\\ contractEnd\end{tabular}} & \multicolumn{1}{c|}{C} & \multicolumn{1}{c|}{\begin{tabular}[c]{@{}c@{}}before and after a new contract \\ is executed \end{tabular}} \\ \hline

\end{tabular}
}

\end{table}

	\smallskip \noindent \textbf{Instrumentation points}\tab
	Our system supports three types of instrumentation points, i.e., \textit{opcode-}, \textit{transaction-} and \textit{contract-oriented} ones.
	Table~\ref{tab:imp_hookpoints} shows an overview of these instrumentation points.
	
	First, the \textit{opcode-oriented} instrumentation links with two callback functions for each opcode, \code{\{op\}} and after
	\code{\{op\}}. They are launched before and after executing the opcode \code{\{op\}}.
	
	Second, the \textit{transaction-oriented} callbacks, including \code{transactionStart} and \code{transactionEnd}, are launched before and after the execution of
	a {normal transaction}. These two instrumentation points are usually used for the initialization and processing results in the
	analysis script. Note that, this type of instrumentation points only works for normal transactions, which are initialized from EOAs.
	For internal transactions that are initialized from smart contracts, they are covered in the \textit{contract-oriented} instrumentation 
	point.
	
	Third, the \textit{contract-oriented} callback functions, including \code{contractStart} and \code{contractEnd}, deal with function calls
	crossing smart contracts (internal transactions). These two functions are invoked at the start and at the end of the execution
	of a smart contract function.

\begin{figure}[t]
	\centering
	\includegraphics[width=0.5\textwidth]{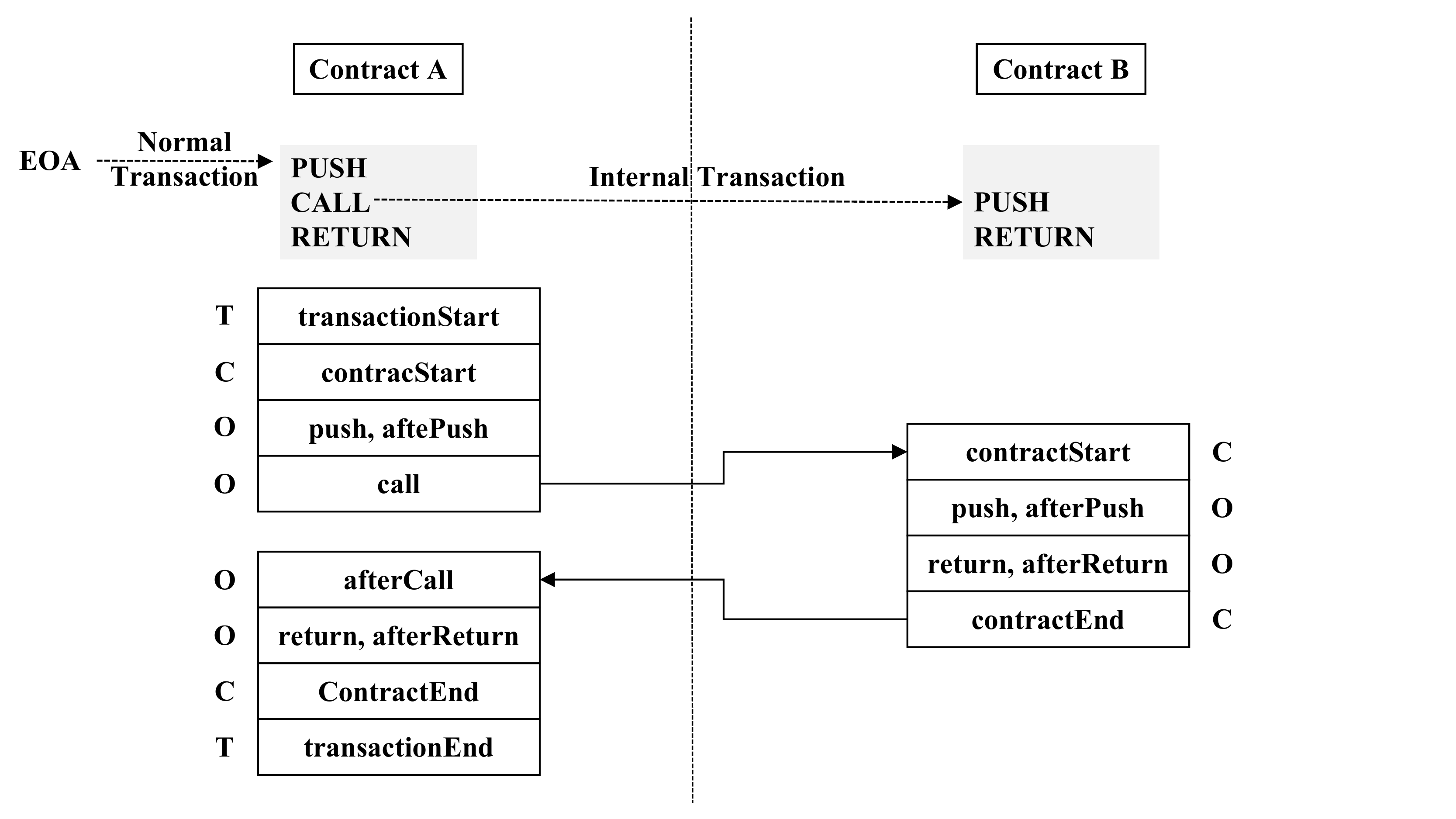}
	\caption{The sequence of invoking callback functions at different types of instrumentation
		points. The code of the smart contract is for illustration only.
		O: \textit{opcode-oriented}; T: \textit{transaction-oriented}; C: \textit{contract-oriented}.}
	\label{fig:imp_hookpoints}
\end{figure}

	Fig.~\ref{fig:imp_hookpoints} shows the sequence of invoking callback functions at different instrumentation points.
	When an EOA issues a normal transaction, \code{transactionStart} will be invoked, and then \code{contractStart} is
	executed. That's because the normal transaction initializes the execution of smart contract A. Then the callback functions
	for each opcode are launched, until the \code{CALL} opcode. This opcode invokes the function inside the smart contract B
	and creates the internal transaction. Since the smart contract B is executed, \code{contractStart} will be invoked again.
	After that, callback functions for different opcodes will be invoked accordingly.
	
	Note that, the execution context is switched from the EVM to the Duktape JavaScript engine, only when a callback function is defined and the instrumentation point is hit at runtime. This minimizes the number of context switches between EVM and Duktape. Compared with the JSTracer inside the Geth, our implementation is more efficient (Section~\ref{subsubsec:eva_engine}).

	\begin{table*}[t]
    \centering
    \caption{APIs provided by our instrumentation framework.}
    \label{tab:imp_hookapis}
    \footnotesize
    \resizebox{0.9\textwidth}{!}{
    \begin{tabular}{|l|l|l|l|l|l|l|}
    \hline
    \multicolumn{7}{|c|}{APIs to retrieve execution context} \\ 
    \hline
    op.getN() & stack.length() & memory.slice(start, end) & contract.getSelfAddress() & getBalance(addr) & getBlockNumber() & getPc() \\
    op.toNumber() & stack.peek(n) & memory.getUint(offset) & contract.getCodeAddress() & getNonce(addr) & getTxnIndex() & getGas() \\
    op.toString() & & & contract.getValue() & getCode(addr) & getTxnHash() & getDepth() \\
    &&& contract.getInput() & getStorage(addr) & & getReturnData() \\
    \hline
    \multicolumn{7}{|c|}{Other APIs} \\
    \hline
    \multicolumn{3}{|c}{cfg.hijack(isJump)} & \multicolumn{4}{|c|}{params.get(key)} \\
    \end{tabular}}
    \resizebox{0.9\textwidth}{!}{
    \begin{tabular}{|l|l|l|l|l|}
    \hline
    \multicolumn{5}{|c|}{APIs to assign, clear and check taint tags} \\
    \hline
    labelStack(n,tag) & labelMemory(offset,size,tag) & labelInput(o,s,t) & labelReturnData(o,s,t) & labelStorage(addr,slot,tag) \\
    clearStack(n) & clearMemory(offset,size) & clearInput(o,s) & clearReturnData(o,s) & clearStorage(addr,slot) \\
    peekStack(n) & peekMemory(offset) & peekInput(o) & peekReturnData(o) & peekStorage(addr,slot) \\
    & peekMemorySlice(offset,size) & peekInputSlice(o,s) & peekReturnDataSlice(o,s) & \\
    \hline
    \end{tabular}}
\end{table*}

	\smallskip \noindent \textbf{APIs to retrieve the execution context}\tab
	Our system provides multiple APIs to get the information of current execution context. Table~\ref{tab:imp_hookapis} shows an overview of these APIs. We elaborate some of them in the following.
	
	\begin{itemize}[leftmargin=*]
		
		\item \textit{Normal transactions.} Attributes of normal transactions are obtained by invoking \code{getBlockNumber}, 
		\code{getTxnIndex} and \code{getTxnHash}. These attributes are used to distinguish different
		normal transactions.
		
		\item \textit{Internal transactions.} Two APIs \code{contract.getSelfAddress}
		and \code{contract.getCodeAddress} are used to retrieve the
		context contract and code contract. The code contract is the address of the callee smart contract. However, the context contract can be the caller and the callee smart contract, depending on the opcode used to invoke the contract. This complies with the definition in Geth~\cite{go-ethereum-contract}.
		The API \code{contract.getValue} returns
		the amount of Ether that is transferred into the code contract.

		Every time an internal transaction starts, the EVM stack depth will be increased by one.
		On the contrary, every time an internal transaction ends, it will be decreased by one.
		The API \code{getDepth} is provided to get current EVM stack depth.
		By using this information, we can detect the occurrence of a recursive function call.
		
		\item \textit{Parameters and return values.}
		The API \code{contract.getInput} returns the input data (parameters) when invoking
		a function, while \code{getReturnData} obtains return values.
		
		\item \textit{The program counter and remaining gas.}
		APIs \code{getPc} and \code{getGas} return the current program counter and remaining gas.
		
		\item \textit{Accounts.}
		APIs \code{getBalance}, \code{getCode}, \code{getStorage} return the current states of an account at any time.
		
	\end{itemize}
	
	\smallskip \noindent \textbf{Dynamic taint engine}\tab
	Dynamic taint analysis has been widely used for security applications. Our framework implements a dynamic taint engine
	that facilitates the development of analysis scripts. 
	
	Our taint analysis engine supports the taint tag propagation crossing different smart contracts. When the EVM triggers
	an internal transaction, it will pass input values from the caller's memory to the callee's input field. When the invocation
	returns, the return value is put into the caller's ret field. We propagate the taint tags in opcodes \code{CALLDATALOAD},
	\code{CALLDATACOPY} and \code{RETURNDATACOPY} that operate stack, memory, ret and input field.
	Table~\ref{tab:imp_hookapis} summarizes APIs to assign, clear and check taint tags. APIs \code{label*} allow an analyst
	to assign taint tags. APIs \code{peek*} and \code{clear*} allow an analyst to check and clear tags.

	\begin{figure}[t]
    \begin{lstlisting}[captionpos=b,basicstyle=\scriptsize,label=list:owned, language=JavaScript]
    {
        sload: function(log){
            contextContract = toHex(log.contract.getSelfAddress())
            key = log.stack.peek(0).toString(16)
            tag = contract+"_"+key
            log.taint.labelStack(0, tag)
        },
        
        jumpi: function(log) {
            tags = log.taint.peekStack(1)
            for (tag in tags) {
                contextContract = tag.substring(0, tag.indexOf("_"))
                key = tag.substring(tag.indexOf("_"))
                console.log("Storage", key, "in contract", contextContract, "influenced the control flow.")
            }
        }
    }
    \end{lstlisting}
    \caption{An example of how to use the dynamic taint engine to assign and check taint tags. }
    \label{fig:imp_taint}
    \vspace{-0.2in}
    \end{figure}
	
	Fig.~\ref{fig:imp_taint} shows an
	example of how to use these APIs. Specifically, two callback functions \code{sload} and \code{jumpi} are invoked before 
	executing opcodes \code{SLOAD} and \code{JUMPI}, respectively. Inside the callback function \code{sload}, it assigns the
	taint tag to the value on the top of the stack (index $0$) using \code{log.taint.labelStack(0, tag)}. Then the taint engine
	will propagate the tag, even crossing different contracts. When the callback function \code{jumpi} is executed, the 
	\code{log.taint.peekStack(1)} checks whether the second value on the stack (index $1$) has the taint tag. If so, it changes the program counter.
	Thus, by checking the taint tag, an analyst can get the storage variables that can influence the control flow.
	
	\section{Evaluation}
	\label{sec:evaluation}
	In this section, we will present the evaluation result of \system{} by answering the following research questions. If not otherwise specified, the evaluation is performed on the dataset that contains the Ethereum state from the genesis block (mined on July 30th, 2015) to the $10,400,000$th one (mined on July 5th, 2020). 
	
	\begin{itemize} [leftmargin=*]
		
		\item\textbf{R1} What's the performance of \system{} and whether \system{} solves the scalability issue?
		
		\item\textbf{R2} Whether \system{} can help understand the behaviors of suspicious transactions and detect more attack instances?
		
		\item\textbf{R3} Whether \system{} performs better than previous systems in terms of detected attacks?
		
	\end{itemize}
	
	To answer \textbf{R1}, we report the comparison result of the storage consumption and the time used to replay transactions. The result shows that \system{} consumes less storage and has a speedup of around $2300$x when replying transactions. This demonstrates the capability of our system to perform the analysis on a large number of transactions.

	To answer \textbf{R2}, we use three different types of public information as inputs, including \textit{a victim smart contract}, \textit{a reported suspicious transaction}, and the \textit{abnormal blockchain state}. For each type of information, our system first understands attack behaviors, and then detect more attack instances. We report the result in Section~\ref{subsec:eva_victim_contract}, Section~\ref{subsec:eva_reported_maltx_as_input}, and Section~\ref{subsec:abnormal_creation_suicidal} (in Appendix), respectively.
	
	To answer \textbf{R3}, we compare the detection result of the re-entrancy attack with previous systems. Our evaluation shows that our system is more accurate than previous ones. We report the result in Section~\ref{subsec:eva_comparison}.

	\subsection{Performance and Scalability}
	\label{subsec:eva_performance}
	In this section, we demonstrate the scalability of our system via evaluating its performance from the following perspectives. First, the storage use is more efficient than previous systems, while at the same it can support the replay of arbitrary transactions. Second, the \aggregator{} can help locate suspicious and candidate transactions in an efficient way. Third, the \engine{} can replay arbitrary transactions, with a 2,300x speedup.
	All experiments were performed on a machine with four CPUs (Intel(R) Xeon(R) Silver $4110$ CPU @ 2.10GHz) and 128GB memory. 
	
	\begin{table}
    \centering
    \caption{The comparison of the storage usage.}
    \label{tab:eva_storage_comparison}
    \footnotesize
    \resizebox{0.5\textwidth}{!}{
    \begin{tabular}{|l|l|l|}
    \hline
     & Blocks & Storage \\
    \hline
    Geth Archive Node & $0~~~~~~~~~~~~$ - $7,635,000$ & $2,320$GB \\
    \hline
    Trace DB of \txspector{} & $0~~~~~~~~~~~~$ - $7,200,000$ & $1,577$GB \\
    \hline
    Logic Relation DB of \txspector{} & $7,000,000$ - $7,200,000$ & $2,949$GB \\
    \hline
    \hline
    \aggregator{} & $0~~~~~~~~~~~~$ - $\textbf{10,507,977}$ &  $\textbf{1,844}$GB \\
    \hline
    \end{tabular}}
\vspace{-0.2in}
\end{table}

	\subsubsection{Storage Use}
	\label{subsubsec:eva_dataaggregator}
	The \aggregator{} in our system stores the saved Ethereum state. 
	We compare the storage use of our system with other ones that also store the Ethereum state. Specifically, ECFCHecker, Sereum, and SODA leverage the \textit{archive node of Geth} to perform the analysis. \txspector{} replays historical transactions in Ethereum to record EVM bytecode-level traces into a \textit{trace DB}, and stores the logic relations into a \textit{logic relation DB}~\cite{TXSPECTOR}.

	As shown in Table~\ref{tab:eva_storage_comparison}, the Geth archive node~\cite{archivedata} of the
	first $7.635$ million blocks uses $2,320$ GB~\footnote{This data is obtained from the official Ethereum  blog~\cite{geth-blog}.}. The trace DB of \txspector{} and logic relation DB of \txspector{} consume $1,577$ GB, and $2,949$ GB for $7.2$ million and $0.2$ million blocks, respectively. Obviously, \txspector{} requires more space to support its analysis.

	Our system costs only $1,844$ GB after collecting the Ethereum state for $10.5$ million blocks. That's because it only collects necessary state information to perform the security analysis and replay transactions. Note that, our system does not scarify the analysis capability to save storage. In particular, even though it consumes less storage, it can fully support the query to locate candidate transactions and replay them, as shown in the experiments to answer \textbf{R2}. This result shows \system{} does not suffer from the scalability issue due to the storage consumption.
	
	\subsubsection{Query Transactions}
	\label{subsubsec:eva_query}
	The \aggregator{} provides an interface to locate transactions by querying the saved Ethereum state, e.g, a normal transaction with more than $1,000$ internal transactions whose Ether transferred are large than a certain amount. Our evaluation shows that most querying tasks can be finished in seconds, while complicated ones may last for a few minutes.
	For instance, the collection of candidate transactions for the re-entrancy attack (Section~\ref{subsec:eva_reported_maltx_as_input}) and the bad randomness attack (Section~\ref{subsec:eva_victim_contract}) both take less than $5$ minutes (retrieved $209,227$ and $10,296,519$ candidates from $754,614,255$ normal transactions) in our experiments.

	\subsubsection{Replay Transactions}
	\label{subsubsec:eva_engine}
	In the following, we will compare the performance of our system with \textit{JSTracer} (in the archive mode) supported by Geth~\cite{go-ethereum}. To the best of our knowledge, this is the only comparable counterpart that can \textit{repeatedly replay and instrument transactions}.
	
	First, we randomly pick $100$ normal transactions that have triggered internal transactions.
	Then, we develop a script that has an equivalent functionality with
	the example~\cite{tracer-example} (\textit{4byte\_tracer.js}) provided by Geth.
	Finally, we use the JSTracer and our system to \textit{replay} $100$ normal
	transactions. Note that, a normal transaction could trigger multiple
	internal transactions, thereby the total number of replayed transactions is $2,519$.
	
	\begin{table}
    \centering
    \caption{The comparison of JSTracer and our system to replay $100$ normal transactions.}
    \label{tab:eva_inst_performance}
    \resizebox{0.5\textwidth}{!}{
    \begin{tabular}{|l|c|c|c|c|}
    \hline
    \multicolumn{2}{|c|}{Tools} & \multicolumn{1}{c|}{\begin{tabular}[c]{@{}c@{}}\textbf{Retrieve}\\\textbf{State}\end{tabular}} &
    \multicolumn{1}{c|}{\begin{tabular}[c]{@{}c@{}}\textbf{Execute}\\\textbf{Script}\end{tabular}} &
    \textbf{Other} \\
    \hline
    \multirow{2}{*}{JSTracer} &   & 39m 6s 997ms & 0m 16s 984ms & 8m 13s 467ms \\
    \cline{2-5}
     & Total & \multicolumn{3}{c|}{\textbf{47m 37s 448ms}} \\
    \hline
    \multirow{2}{*}{Our system} &   & 0m 0s 446ms & 0m 0s 217ms  & 0m 0s 544ms \\
    \cline{2-5}
    & Total & \multicolumn{3}{c|}{\textbf{0m 1s 207ms}} \\
    \hline
    \end{tabular}}
\end{table}

	Table~\ref{tab:eva_inst_performance} shows the comparison result of the transaction replay time between JSTracer and our system. 
	Specifically, JSTracer spends more than $47$ minutes to replay the transactions, while our system takes only around one second to replay them. The result suggests that our system outperforms JSTracer with an around $2,300$x speedup. We further explore the possible reasons.
	
	\begin{itemize} [leftmargin=*]
		\item{\textbf{Granularity of the Ethereum historical state}}\tab To retrieve the Ethereum historical state of the $100$ normal transactions, JSTracer had to \textit{replay} $3,289$ additional normal transactions.
		However, \system{} can directly query fine-grained accounts' state information from the \aggregator{}.
		\smallskip
		\item{\textbf{Number of context switches}}\tab JSTracer needs to switch to the JavaScript environment for every opcode. Alternatively, our \framework{} only performs context-switch when instrumentation points are hit. That is why JSTracer performed $1,305,864$ context switches, while \system{} only performed $2,502$ ones.
	\end{itemize}

	The result demonstrates that our system can replay a large number of transactions. 
	In fact, for the $10,296,519$ normal transactions used to detect the new instances of the bad randomness
	attack, our system took {12 hours 7 minutes} to replay all of them, which is quite difficult (if not impossible) for other systems to complete such a task. 
	
	\smallskip \noindent
	\textbf{Answers to Q1:}\tab Our system consumes less storage than other systems, while at the same time the stored Ethereum state can support replaying arbitrary transactions. 
	Besides, the \engine{} in our system is more efficient when replaying transactions. The lower storage consumption and efficient replay engine make the detection of attacks in the whole Ethereum blocks possible.

	\subsection{Type-I Input: A Victim Contract}
	\label{subsec:eva_victim_contract}
	
	An analyst may receive incomplete information, e.g., a smart contract is being attacked. However, there is no detailed information about the vulnerability of the victim contract, nor the information on how the attack works. Our system can help an analyst understand the attack, and detect more attack instances.
	We use the Fomo3D~\cite{airdrop} as an example to illustrate how \system{} helps analysts reveal
	attacks from a victim smart contract. The input to our system is the address~\footnote{\scriptsize{0xa62142888aba8370742be823c1782d17a0389da1}} of the victim smart contract.

	\subsubsection{Understand the Attack}
	As shown in Fig.~\ref{fig:investigation}, an analyst leverages our system to understand the attack behaviors. 
	
	\begin{figure}[t]
	\centering
	\includegraphics[width=0.6\textwidth]{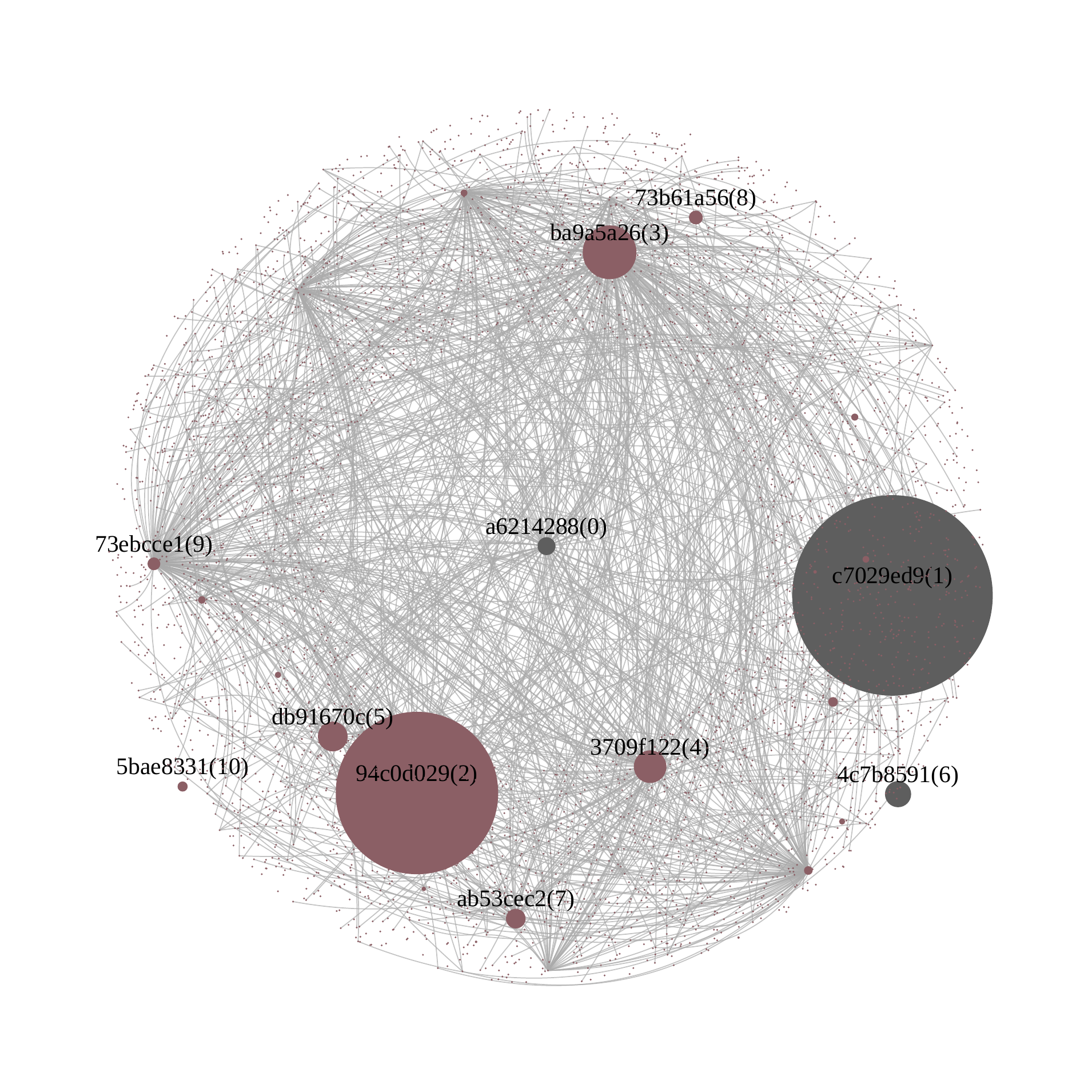}
	\caption{The money flow graph of the Fomo3D smart contract. For better
		illustration, we use $180,244$ transactions to generate this graph.
		The total number of transactions with Fomo3D is much larger. }
	\label{fig:eva_moneyflow}
	\vspace{-0.1in}
\end{figure}

	\smallskip \noindent \textbf{Locate suspicious transactions}\tab
	To locate suspicious transactions that may involve in the attack, 
	our first step is to construct the \textit{money flow graph} to locate suspicious accounts.
	That's because the Fomo3D is a gambling app. The money will flow into (successful) attackers (and other lucky players).
	Fig.~\ref{fig:eva_moneyflow} shows the money flow graph
	constructed using transactions retrieved from the \aggregator{}.
	Specifically, nodes in the graph represent accounts, 
	and edges represent the direct and indirect transactions with
	the Fomo3D game. The size of each node denotes the number of
	Ether it receives.
	
	We observe that several accounts have a much larger size
	than others. It means these accounts have received much more Ether from the
	game than others. Initial analysis shows that three of them belong to
	Fomo3D (number $0$, $1$, and $6$). We then take further analysis for
	other accounts. 
	
	\begin{figure}[t]
	\centering
	\includegraphics[width=0.6\textwidth]{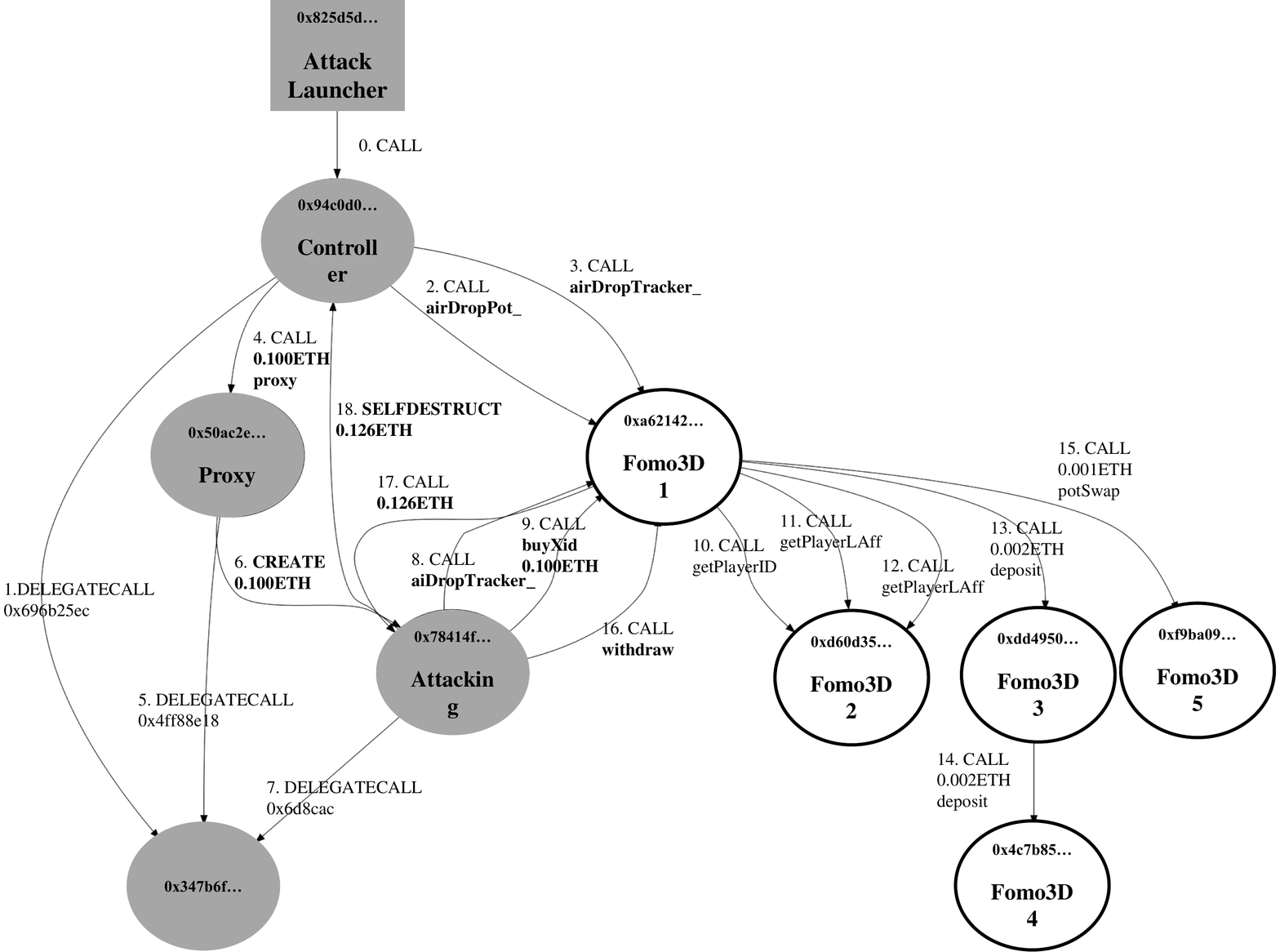}
	\caption{The dynamic call graph of a suspicious transaction. We draw three types of information for an internal transaction: 1. Serial number and the opcode to trigger an internal transaction; 2. Transferred Ether, null means no Ether transferred; 3. Invoked 
	function (we search the name from the 4byte function signature database~\cite{4byte}), null means that the input data is empty;
	(Square: EOA, Circle: smart contract; Grey Box: attacker, White Box: victim.)}
	\label{fig:eva_fomo3dattack_callgraph}
\end{figure}

	\smallskip \noindent \textbf{Understand suspicious transactions}\tab
	We analyze a normal transaction~\footnote{\scriptsize{0xee95751e94c8427f94ddf34e15bb322f681a0d264e9d2d21c3fc0d687dff22c2}} that invokes the 
	smart contract (index $2$ in the money flow graph)~\footnote{\scriptsize{0x94c0d029a7b64bf443e89c5006089364c0d60d61}} to receive
	Ether from Fomo3D. To this end, we construct the dynamic call graph in Fig.~\ref{fig:eva_fomo3dattack_callgraph}. 
	The nodes in the graph represent accounts (both EOA and smart contracts), and the edges denote Ether transfer or function invocation.

	The
	call sequence of this graph shows that, the contract (\code{0x94c0d0}) transfers
	$0.1$ Ether to the contract (\code{0x50ac2e}) (index $4$), which further creates
	a new smart contract (\code{0x78414f}) (index $6$). This \textit{new} contract
	buys the key (index $9$) with $0.1$ Ether and then receives $0.126$ Ether
	(index $17$) from the game. The received Ether is transferred back to the contract
	(\code{0x94c0d0}) with a \code{SELFDESTRUCT} operation (index $18$).
	During this process, it obtains a profit of $0.026$ Ether.

	There also exist many similar transactions related to the contract (\code{0x94c0d0}). These transactions get a lot of rewards from the Fomo3D game.
	We suspect the contract (\code{0x94c0d0}) has a mechanism to predict 
	whether it can win the bonus before playing the game. Otherwise, it can not win every time.
	After locating all the transactions and smart contracts created from this account by querying
	the \aggregator{}, we find that the contract (\code{0x94c0d0}) indeed can predict whether it can win.
	That's because the Fomo3D game uses the address of the player (controlled by the attacker) as one of the sources to generate the random number that determines the winner.
	
	\begin{figure}[t]
	\centering
	\includegraphics[width=0.5\textwidth]{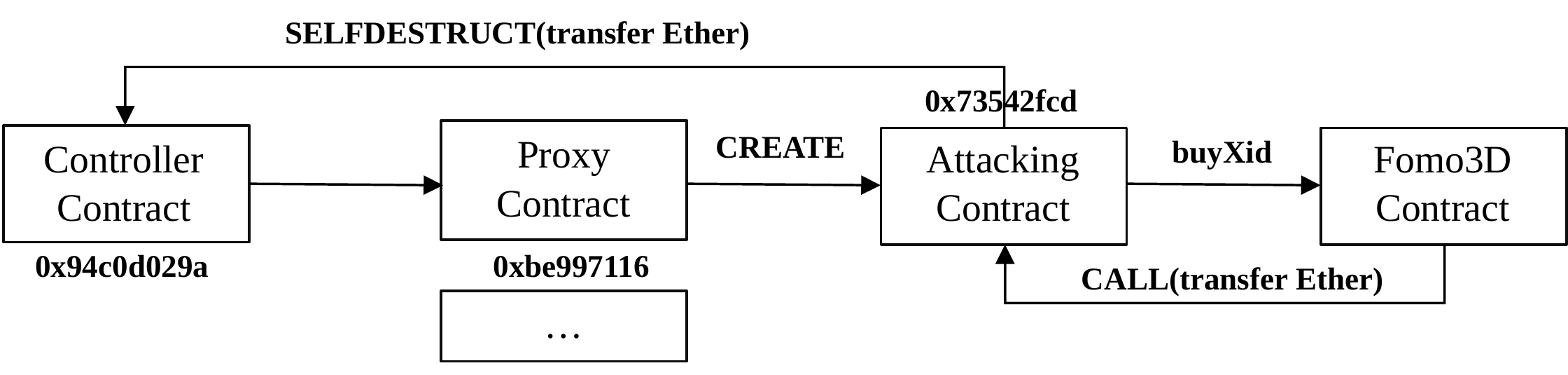}
	\caption{The flow of the bad randomness attack.}
	\label{fig:eva_fomo3dattackflow}
\end{figure}
	
	Fig.~\ref{fig:eva_fomo3dattackflow} shows (a simplified version of) the attack flow.
	There is a controller contract, which creates a lot of proxy contracts
	(more than $1,000$) in advance.
	Then during the attack, the controller attack loops through each proxy
	contract. It calculates the address of a newly created smart contract (but does not create it.) because the address is predictable (Section~\ref{subsec:b_accounts}).
	Then it uses this address and the block information to predict
	whether it will get the bonus by executing the same logic with the Fomo3D game. 
	If so, the proxy smart contract creates
	the attacking contract, which further buys the key to play the game and
	win the bonus. After that, the attacking contract self-destructs itself
	to transfer the earned bonus to the controller smart contract.
	
	Because the attack exploits the vulnerable process of the smart contract to generate a random number, we name this attack as the \textit{bad randomness attack}.
	
	\subsubsection{Detect more attack instances}
	After understanding the above attack, we then use our system to detect more bad randomness attacks. Specifically, we first use the \aggregator{} to filter out transactions that are not related to the attack. Then we use the \engine{} to replay the remaining transactions and the \framework{} to confirm new attack instances at runtime.

	\smallskip \noindent \textbf{Locate candidate transactions}\tab
	In order to avoid replaying unnecessary transactions (costing lots of time), we first use
	the \aggregator{} to remove transactions that are not related to the bad randomness attack.
	
	We label normal transactions that fulfill the following requirements as candidate transactions. First, it has triggered more than one internal transaction. Second, the triggered internal transaction has transferred Ether to another smart contract. That's because in order to launch the attack, attackers have to use a contract to transfer Ether to play the game, thus creating an
	internal transaction. This rule is conservative. It may label some benign transactions as candidates. However, we want to include as many candidate transactions as possible in this step and leverage
	the \engine{} to confirm whether they are real attacks. In total, our system locates $10,296,519$ candidate transactions.

	\smallskip \noindent \textbf{Confirm the bad randomness attack}\tab
	After locating the candidate transactions, we then use the 
	\engine{} to replay them and confirm attacks at runtime.
	
	The key observation of this attack is that the malicious contract is
	using the same algorithm to generate the random number as the victim contract. We develop the detection script as follows.

	\begin{enumerate}[leftmargin=*]
		\item First, we find all the variables that are generated from block information, e.g., coinbase, gaslimit and etc. This is implemented using our taint analysis engine by setting the block information as taint sources.
		\item Second, for each variable \textit{v} found in the previous step, we check whether it influences the control flow of the smart contract. That's because we only care about the variables that can determine the winner. If so, we log its execution context \textit{C}.
		\item If there exist two same execution contexts in different internal transactions that are triggered by a same normal transaction, then the normal transaction is a malicious one that launches the attack. That's because two smart contracts are executing the same algorithm that uses the same random number sources to generate a variable that can influence the control flow to determine the winner.
	\end{enumerate}
	
	\smallskip \noindent \textbf{Detection result}\tab
	We replayed $10,296,519$ candidate transactions with our analysis script.
	After that, $40,449$ normal transactions are labeled as malicious ones.
	During this process, $272$ malicious smart contracts are detected.
	We then group them based on their creators, i.e.,
	EOAs that create these contracts. In
	total, we get $79$ groups. We manually checked the malicious smart contracts created in each
	group and found that $74$ of them are true positives.
	In total, they have initialized $40,358$ normal transactions
	to attack $95$ victim smart contracts, which includes various gambling games.
	Table-I in the link~\footnote{https://github.com/Anonymouspaper146/SP2021fallsubmission} shows
	the detailed information of victim contracts and the false positives.

	\subsection{Type-II Input: A Reported Suspicious Transaction}
	\label{subsec:eva_reported_maltx_as_input}
	Besides the victim contract, an analyst may receive the information that a malicious transaction is attacking a smart contract. Though there may exist partial information of the attack, the details of the attack are unknown. 
	
	\subsubsection{Understand the Attack}
	Attackers leveraged the re-entrancy vulnerability to launch the attack towards the DAO smart contract and stole $3.6$ million Ether~\cite{dao}. 
	In this following, we will elaborate on the process to understand the attack by leveraging a reported transaction~\footnote{\scriptsize{0xfb6526b62f0a4627543cba59a24b9790d0f53ecd841b0adc6ba0026cadf77715}}. Then we will leverage the gained knowledge to detect more re-entrancy attacks.

	\begin{figure}[t]
	\centering
	\includegraphics[width=0.6\textwidth]{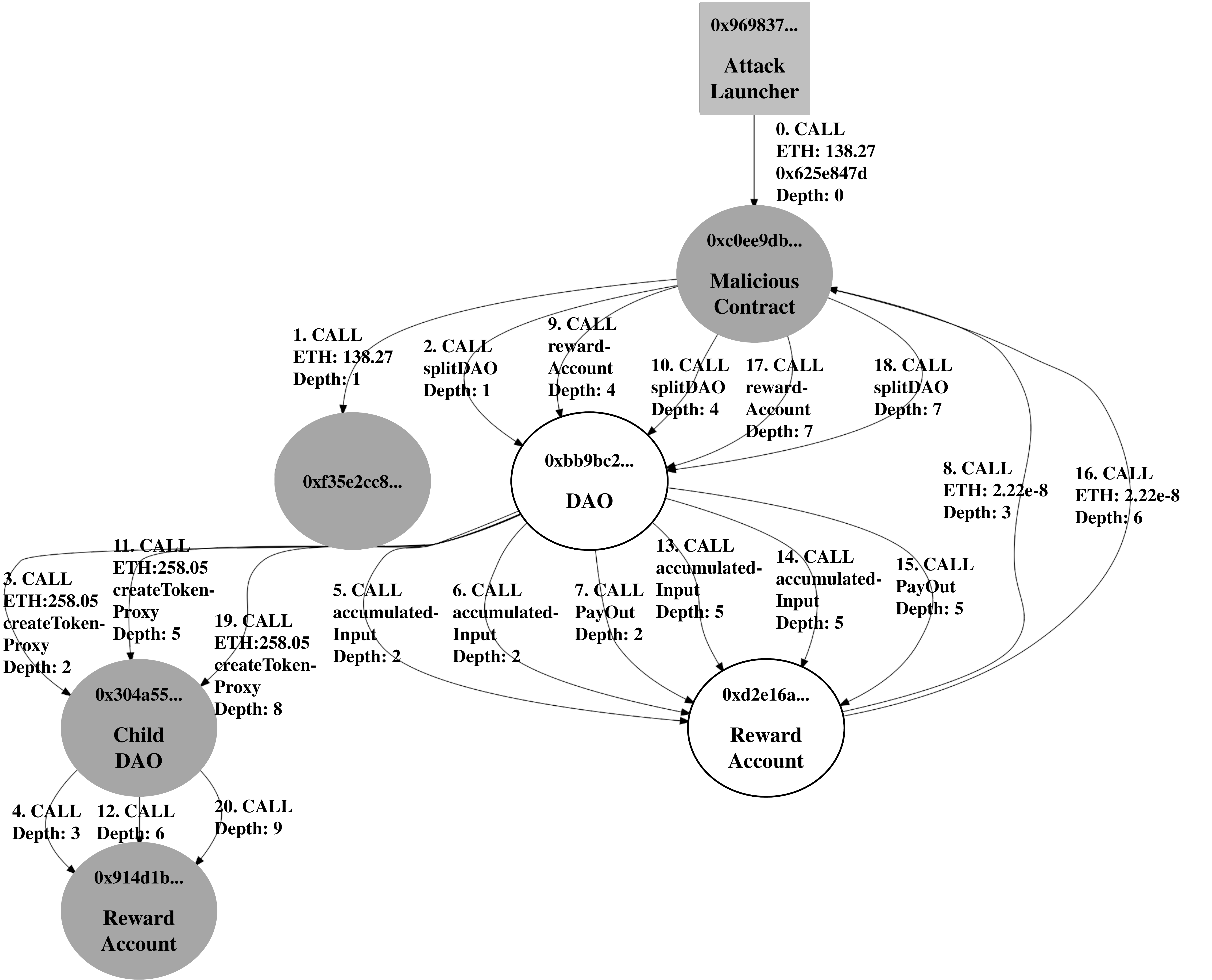}
	\caption{The dynamic call graph of a suspicious transaction that exploits the DAO smart contract. We use four lines to describe an internal transaction: 1. Serial number and the opcode to trigger an internal transaction; 2. Transferred Ether, null means no Ether transferred; 3. Invoked 
		function (we search the name from the 4byte function signature database~\cite{4byte}), null means that the input data is empty; 4. EVM stack depth.
		({Square: EOA, Circle: smart contract; Grey Box: attacker, White Box: victim;})}
	\label{fig:eva_daoattack_call_graph}
\end{figure}
	
	\smallskip \noindent \textbf{Understand suspicious transactions}\tab
	The input is a reported transaction, e.g., from a public forum. An analyst needs to understand how the attack works. 
	
	We construct a dynamic call graph in Fig.\ref{fig:eva_daoattack_call_graph}.
	The serial numbers of transactions are in chronological order. The $0$th transaction is a normal transaction, and others
	are internal transactions triggered by the normal transaction. For better illustration, we only use the first $20$ internal
	transactions to draw the graph. The actual number of internal transactions is $185$.

	By analyzing this graph, we can find two distinct features of transactions that launch the attack. First, there  
	exists a loop in the graph. This is reasonable since the call to the fallback function that further invokes the vulnerable
	contracts will create a loop in the call graph. For instance, internal transactions $2$, $7$, and $8$ create a loop that starts
	from and ends at the malicious contract (\code{0xc0ee9db}). Second, there should exist a special smart contract called 
	\textit{reentry point}, which is the smart contract that will be invoked again before its previous invocation completes. For
	instance, the DAO contract (\code{0xbb9bc2}) is a \textit{reentry point}, since the EVM stack depths of internal transactions
	(index $3$ to $9$) are all bigger than internal transaction $2$. That means before an invocation to the DAO contract \code{0xbb9bc2}
	(internal transaction $2$) returns, another invocation (internal transaction $9$) to the same contract happens.
	
	\subsubsection{Detect more attack instances}
	After understanding the re-entrancy attack, we detect more attack instances.
	
	\smallskip \noindent \textbf{Locate candidate transactions}\tab
	According to the gained knowledge of the attack in the previous step, we use the following two rules to locate candidate transactions. We label a normal transaction as a candidate when it satisfies the following two conditions.
	
	\begin{enumerate}[leftmargin=*]
		\item First, internal transactions triggered by this normal transaction create
		a loop that contains at least one \textit{reentry point}. This detects the existence
		of reentrant function calls.
		\item Second, there is \textit{at least one internal transaction} that involves with the Ether
		or ERC20 token transfer. This rule is to remove transactions that do not cause any
		change to the Ether or ERC20 tokens. They are not real attacks since no financial benefits
		are achieved during this process. 
	\end{enumerate}
	
	Thanks to the query interface provided by the \aggregator{}, we can easily locate candidate transactions and remove unrelated ones. In total, we get $209,227$ candidate transactions.
	
	\smallskip \noindent \textbf{Confirm the re-entrancy attack}\tab
	We further replay candidate transactions to confirm the re-entrancy attack at runtime.
	During this process, an analysis script
	is invoked. 
	Our system first constructs a set of variables that could influence jump targets of the \code{JUMPI}
	opcode or values of transferred Ether. Thanks to the dynamic taint engine of our system, we can check whether a variable
	could influence the control flow by checking the taint tag of the second top value on the stack (\code{taint.peekStack(1)}).
	For each variable $v$ in this set, we define the callback function for the \code{SSTORE} opcode to monitor whether the
	variable has been updated after the re-entrant point. If so, we will label the normal transaction as malicious. 
	
	\smallskip \noindent \textbf{Detection result}\tab
	\system{} locates $209,227$ candidate transactions.
	After replaying them, our system detected $2,973$ malicious normal transactions in the wild. 
	Attackers are targeting $52$ victim contracts, which are shown in Table-II in the link~\footnote{https://github.com/Anonymouspaper146/SP2021fallsubmission.}.
	
	We manually analyzed each detected attack. During the analysis, we only
	consider transactions that have caused financial loss as true positives
	(real attacks). Our analysis shows that $46$ transactions are false positives, which are related to $4$ 
	victims (marked with * in Table-II).
	We show a detailed analysis of one false positive in the following.
	
	\begin{figure}[t]
	\begin{lstlisting}[captionpos=b,basicstyle=\scriptsize,label=list:owned]
	function doWithdraw(address from, address to, uint256 amount) internal {
	// only use in emergencies!
	// you can only get a little at a time.
	// we will hodl the rest for you.
	
	require(amount <= MAX_WITHDRAWAL);
	require(balances[from] >= amount);
	require(withdrawalCount[from] < 3);
	
	balances[from] = balances[from].sub(amount);
	// reentry point
	to.call.value(amount)();
	withdrawalCount[from] = withdrawalCount[from].add(1);
	}
	\end{lstlisting}
	\caption{The code snippet of HODLWallet.}
	\label{fig:eva_fp_code}
	\vspace{-0.2em}
\end{figure}
	
	Our system reported one attack targeting HODLWallet. However it is a false positive since it does not cause the financial loss.
	Fig.~\ref{fig:eva_fp_code}
	shows the code snippet of the \code{doWithdraw} function. Specifically,
	the variable \code{withdrawalCount[from]} in line $8$ influences the control
	flow. Also, this variable is updated after the reentry point in line $13$. Thus,
	our system detects this as a re-entrancy attack.
	However, the transaction does not cause any financial loss since the balance
	\code{balances[from]} has been updated in line 10 (before the reentry point.)
	This is a false positive, though technically it is still a re-entrancy attack that
	targets \code{withdrawalCount[from]} instead of \code{balances[from]}.
	
	Since the DAO attack, the security community has paid lots of attentions to detect this vulnerability. However, the re-entrancy attack still happened recently. Specifically, our system  
	detected $579$ re-entrancy attacks after the $9,200,000$th block (Jan 2nd, 2020), in which $46$ attacks are targeting Lend.Me~\cite{Lend.Me} and $529$ attacks are targeting Uniswap~\cite{Uniswap}. Both of them are DeFi applications. These two attacks caused significant financial loss.

	\smallskip \noindent
	\textbf{Answers to Q2:}\tab With three different types of inputs, our system can help understand the suspicious transactions and further detect new attacks by locating and replaying candidate transactions. This demonstrates the effectiveness of our system to facilitate the attack investigation and detect new attack instances.

	\begin{table*}[t]
    \centering
    \caption{The comparison between our system and others in detecting the Re-entrancy attacks.}
    \label{tab:eva_effectiveness_comparison}
    \footnotesize
    \resizebox{0.8\textwidth}{!}{
    \begin{tabular}{ccccc}
    \toprule[1.3pt]
    Block Range & \# of Normal Transactions & Tools & \# of Flagged Contracts & \# of True Positives \\
    \hline
    \multirow{2}{*}{$0~~~~~~~~~~~~$ - $3,918,380$} & \multirow{2}{*}{$32,048,852$} & ECFChecker~\cite{ECFChecker} & $9$ & $5$ \\
    & & \system{} & \textbf{6} & \textbf{6} \\
    \toprule[1.3pt]
    &  &  & \# Flagged Normal Transactions &  \\
    \hline
    \multirow{2}{*}{$0~~~~~~~~~~~~$ - $9,000,000$} & \multirow{2}{*}{$590,040,664$} & Sereum~\cite{Sereum} & $245,519$ & - \\
    & & \system{} & \textbf{2,392} & \textbf{2,347} \\
    \toprule[1.3pt]
    &  &  & \# Flagged Contracts &  \\
    \hline
    \multirow{2}{*}{$0~~~~~~~~~~~~$ - $8,180,000$} & \multirow{2}{*}{$500,930,221$} & SODA~\cite{chensoda} & $31$ & $27$ \\
    & & \system{} & \textbf{29} & $27$ \\
    \toprule[1.3pt]
    &  &  & \# Flagged Contracts &  \\
    \hline
    \multirow{2}{*}{$0~~~~~~~~~~~~$ - $4,500,000$} & \multirow{2}{*}{$78,141,322$} & ÆGIS~\cite{ferreira2019aegis} & $7$ & $7$ \\
    & & \system{} & $7$ & $7$ \\
    \toprule[1.3pt]
    &  &  & \# Flagged Contracts &  \\
    \hline
    \multirow{2}{*}{$7,000,000$ - $7,200,000$} & \multirow{2}{*}{$9,661,593$} & \txspector{}~\cite{TXSPECTOR} & $30$ & $0$ \\
    & & \system{} & \textbf{1} & \textbf{1} \\
    \toprule[1.3pt]
    \end{tabular}}
\vspace{-0.2em}
\end{table*}

	\subsection{Comparison with previous systems}
	\label{subsec:eva_comparison}
	In this section, we compare our system with previous ones.
	We use the result of the re-entrancy attack since most systems can detect this attack.
	Table~\ref{tab:eva_effectiveness_comparison} shows the overall result. For each system, we use the same dataset and compare the detected attacks. The result shows that our system has lower false positives and false negatives.

	\smallskip \noindent \textbf{ECFCHecker}\tab
	ECFCHecker~\cite{ECFChecker} reports nine malicious smart contracts before $3,918,380$th block (Jun 23, 2017). Among them, five are true positives and four are false positives. Our system detects six malicious smart contracts. All of them are true positives. Specifically, five false positives are the same smart contracts detected by ECFCHecker. One true positive~\footnote{\scriptsize{0xf01fe1a15673a5209c94121c45e2121fe2903416}} (a malicious smart contract in the $1,743,596$-th block) is missed by ECFCHecker. Besides, our system does not flag the four false positives reported by ECFCHecker.

	\smallskip \noindent \textbf{Sereum}\tab
	Sereum~\cite{Sereum} has released the evaluation result for the first $9$ million blocks on GitHub. It flags $245,519$ normal transactions as re-entrancy attacks. Among the first $9$ million blocks, $2,392$ are detected by our system. Besides, among $2,392$ normal transactions, $12$ are not flagged by Sereum.
	
	First, we manually confirm that these $12$ normal transactions are true positives. That means they have been missed by Sereum. Second, for the $243,139$ normal transactions that are flagged by Sereum, we  randomly pick up $10$ transactions. The manual analysis shows that they are all false positives.

	\smallskip \noindent \textbf{SODA}\tab
	For the first $8.18$ million blocks, SODA~\cite{chensoda} reports $31$ vulnerable contracts, with $5$ false positives and $26$ true positives. After double-checking the $31$ contracts, we find two of them
	are false positives~\footnote{\scriptsize{0x72f60eca0db6811274215694129661151f97982e}, \scriptsize{0xd4cd7c881f5ceece4917d856ce73f510d7d0769e}}
	and one is true positive~\footnote{\scriptsize{0x59abb8006b30d7357869760d21b4965475198d9d}} (reported as the false positive by SODA.)
	Therefore, the result is $27$ true positives and four false positives. \system{} detects the same $27$ true positives. 
	
	\smallskip \noindent \textbf{ÆGIS}\tab
	ÆGIS~\cite{ferreira2019aegis} reports that seven smart contracts are victims of
	the re-entrancy attack during the first $4.5$ million blocks.
	\system{} detects the same victimized smart contracts. However, ÆGIS marks fewer attacks transactions ($1,118$ vs $2,301$) than \system{}. That's because
	ÆGIS limits their analysis to the first $10,000$ normal transactions of each contract to 
	reduce the execution time. Our system does not have this limitation, thanks to the efficient \engine{}.
	
	\smallskip \noindent \textbf{\txspector{}}\tab
	Due to the storage consumption, \txspector{} detects the re-entrancy attack from $7,000,000$th block to $7,200,000$th block. It flags 
	$3,357$ normal transactions as malicious and $30$ vulnerable smart contracts. Among them, they manually labeled $17$ ones 
	as true positives. \system{} flags one malicious normal transaction~\footnote{\scriptsize{0xb5c10dbb51b00199d4d817488490f129e80832a4fd6dbf209277c11d42873cca}}
	and one victim contract~\footnote{\scriptsize{0xf91546835f756da0c10cfa0cda95b15577b84aa7}}. It is the re-entrancy attack to SpankChain~\cite{spankchain}. 
	
	The authors of the  \txspector{} kindly provide their dataset for us.
	We manually analyze the $17$ smart contracts that are reported as true positives by \txspector{}. However,
	they are not vulnerable and cannot be victims of the re-entrancy attack according to our definition (causing a financial loss).
	Moreover, one true positive (the SpankChain re-entrancy attack) reported by our system is not detected by \txspector{}.
	
	\smallskip \noindent
	\textbf{Answers to Q3:}\tab Comparing with previous systems, \system{} has lower false positives and false negatives when detecting the re-entrancy attack.

	\section{Discussion}
	\label{sec:discussion}

	The purpose of our system is to detect real attacks. 
	Compared with other static analysis tools~\cite{MAIAN, Ponzi, Vandal, Osiris, Oyente, ContractFuzzer, VeriSmart, VerX, Securify, Zeus}, our system may miss some vulnerable smart contracts that \textit{are not exploited in the wild}.
	Nevertheless, our system does not intend to replace existing static tools. Instead, these tools are complementary to our system. For instance, the vulnerable smart contracts reported by them~\cite{MAIAN, Ponzi, Vandal, Osiris, Oyente, ContractFuzzer, VeriSmart, VerX, Securify, Zeus} could be one type of inputs (as shown in Section~\ref{subsec:eva_victim_contract}) to locate \textit{real attacks}.

	Though the main usage of our system is to perform investigation on attacks that have happened, it can be extended to conduct real-time detection of attacks. We can continuously monitor the blockchain state and use some heuristics to locate suspicious transactions. For instance, we can continuously monitor the transactions that are involved in big-amount Ether transfer. We can mark them as suspicious and understand the purpose of such transactions using our system. Another example is monitoring the transactions with smart contracts that may potentially be attacked, e.g., DeFi applications. That's because such applications are high-value targets for attackers to make profits.
	We leave the real-time detection of new attacks as one of the future work.

	Though we have demonstrated the effectiveness of
	our system, an analyst still needs some public information as inputs, e.g., victim contracts.
	One potential direction is to use
	new techniques, e.g., machine learning algorithms to automatically
	locate suspicious transactions. Currently, our system provides a
	dynamic taint engine to facilitate the analysis. In the future, we can
	integrate more components, e.g., dynamic symbolic execution, into
	the system to ease the development of analysis scripts

	\section{Related Work}
	\label{sec:related}
	\noindent \textbf{Data analysis frameworks of Ethereum}\tab
	Chen et al.~\cite{GraphAnalysis} proposed a graph-analysis based approach
	to analyze Ethereum from different aspects, including money flow,
	account creation and contract invocation.
	DataEther~\cite{DataEther} first instruments an Ethereum full node to
	collect data and then uses ElasticSearch~\cite{elasticsearch} to
	store the collected data. 
	Similar to \system{}, these systems can be used to locate suspicious transactions. 
	However, they are not capable of introspecting the execution of smart contracts to understand and detect more attacks.

	\smallskip
	\noindent \textbf{Static analysis tools of Ethereum smart contracts}\tab
	A number of static analysis tools have been proposed to detect vulnerabilities of Ethereum smart contracts,
	including Oyente~\cite{Oyente}, Mythril~\cite{mythril},
	Osiris~\cite{Osiris}, MAIAN~\cite{MAIAN},
	ContractFuzzer~\cite{ContractFuzzer}, ILF Fuzzer~\cite{he2019learning},
	Securify~\cite{Securify} and ZEUS~\cite{Zeus}.
	These systems only provide a \textit{static} view of
	smart contracts, i.e., whether they are vulnerable or not.
	They cannot provide a \textit{dynamic} view of contract interactions (or transactions), which is useful to analyze and understand attacks. 
	Our system does not intend to replace existing static tools. Instead, they are complementary to our system. For instance, the vulnerable smart contracts reported could be one type of inputs (as shown in Section~\ref{subsec:eva_victim_contract}) to locate \textit{real attacks}.

	\smallskip
	\noindent \textbf{Dynamic analysis tools of Ethereum smart contracts}\tab
	Dynamic analysis has been regarded as an effective complement to static analysis for security purposes. 
	ECFChecker~\cite{ECFChecker}, Sereum~\cite{Sereum}, SODA~\cite{chensoda} and
	ÆGIS~\cite{ferreira2019aegis} are representative tools to analyze Ethereum smart contracts.
	On one side, both Sereum~\cite{Sereum} and ECFChecker~\cite{ECFChecker} focus on the detection of the re-entrancy attack.
	On the other side, SODA~\cite{chensoda} and ÆGIS~\cite{ferreira2019aegis} provide extensible interfaces to detect multiple types of attacks.
	Unfortunately, these tools suffer from the scalability issue. They are not suitable to perform the large-scale detection.

	P{\'{e}}rez et al.~\cite{Datalog} presented the first work that adopts the datalog-based approach to analyze vulnerabilities of smart contracts. However, it only analyzes transactions related to the smart contracts flagged by other tools. 
	\txspector{}~\cite{TXSPECTOR} also relies
	on datalog and supports customized rules to analyze different types of vulnerabilities and attacks.
	However, \txspector{} is not scalable to perform the large-scale detection,  due to the heavy storage consumption.

	Zhou et al.~\cite{everevolvinggame} investigated attacks in the wild.
	They leveraged internal transactions information (named \textit{trace} in the paper) and transaction logs to measure six types of vulnerabilities, including call injection, re-entrancy, integer overflow, airdrop hunting, honeypot, and call-after-destruct. Our system has a different purpose. It focuses on building a scalable framework to understand and detect different types of attacks.
	
	\section{Conclusion}
	\label{sec:conclusion}
	
	In this paper, we present the design of a scalable attack detection framework on Ethereum. It overcomes the scalability issue of existing systems that it can perform timely attack investigation and detect more attacks. We implement a prototype named \system{} and solve three technical challenges. 
	The performance evaluation shows that our system can solve the scalability issue. The result with three different types of information as inputs shows that it can help an analyst understand attack behaviors and further detect more attacks.

	\bibliographystyle{IEEEtranS}
	\bibliography{main}

\begin{thebibliography}{10}
\bibitem{spankchain}
``We got spanked: What we know so far.''
\end{thebibliography}

	\clearpage
	
	\section{Appendix}
	\begin{figure}[t]
    \centering
    \includegraphics[width=0.6\textwidth]{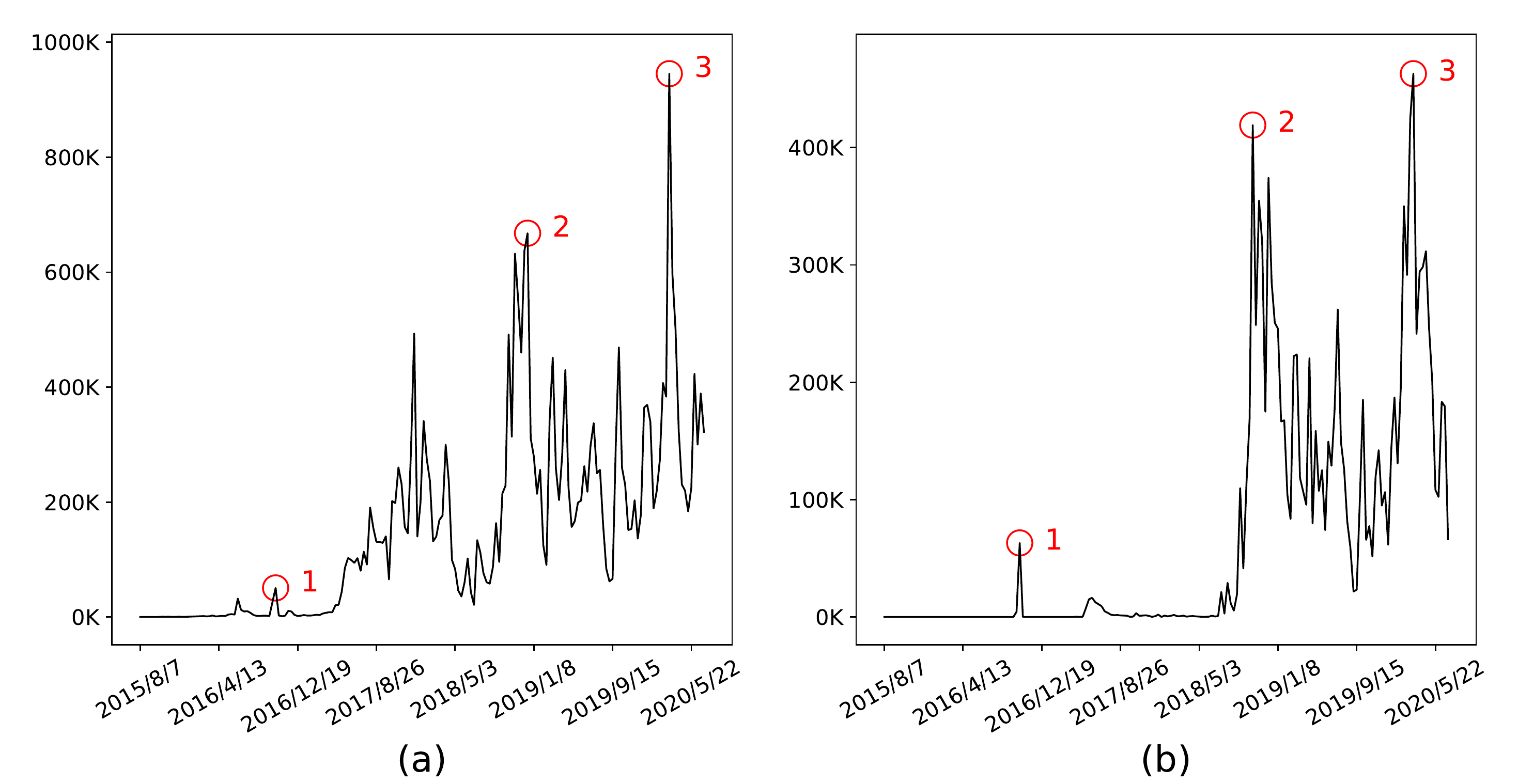}
    \caption{The trend graph of smart contract creation (a) and self-destruction (b). The y-axes show the
    total number of newly created smart contracts and destroyed ones for every ten days, respectively.  }
    \label{fig:eva_creation_and_suicidal_trend}
\end{figure}

	\subsection{Type-III Input: Abnormal Blockchain State}
	\label{subsec:abnormal_creation_suicidal}
	Besides the reported victim smart contracts and malicious transactions, an analyst can leverage the \aggregator{} to observe the blockchain state and use multiple heuristics to locate suspicious transactions. In the following, we elaborate the method of using the number of smart contract creation and self-destruction to locate suspicious transactions, and the process of understanding these transactions to detect multiple types of attacks.

	\subsubsection{Understand the attack}
	Some attacks may lead to abnormal blockchain state, which can be used by an analyst to perform the detection. In the following, we illustrate how our system leverages the abnormal blockchain state to detect attacks.

	\smallskip \noindent \textbf{Locate suspicious transactions}\tab
	Attackers often create malicious smart contracts to automatically launch attacks. After that, they often destroy these contracts to save cost or hide traces. For instance, attackers of the bad randomness attack create a large number of smart contracts to lunch the attack and destruct them afterwards (Section~\ref{subsec:eva_victim_contract}).

	Inspired by this observation, we draw a trend graph of smart contract creation and self-destruction shown in Fig.~\ref{fig:eva_creation_and_suicidal_trend}. 
	From the figure, we can find that there exist several abnormal points where the numbers of 
	new smart contracts (and destroyed ones) are much larger than those of the neighbors (marked with red circles in the figure).

	These three abnormal points appear in blocks ranging from $2,000,000$th to $3,000,000$th,
	$6,500,000$th to $7,500,000$th and $8,900,000$th to $10,110,000$th, respectively. 
	We use \aggregator{} to lookup transactions and accounts that create or destroy these smart contracts and label them as suspicious.

	\smallskip \noindent \textbf{Understand suspicious transactions}\tab
	After analyzing suspicious transactions, we observe two types of attacks and an automated arbitrage trading behavior. We illustrate them in the following.

	\begin{figure}[t]
    \centering
    \includegraphics[width=0.6\textwidth]{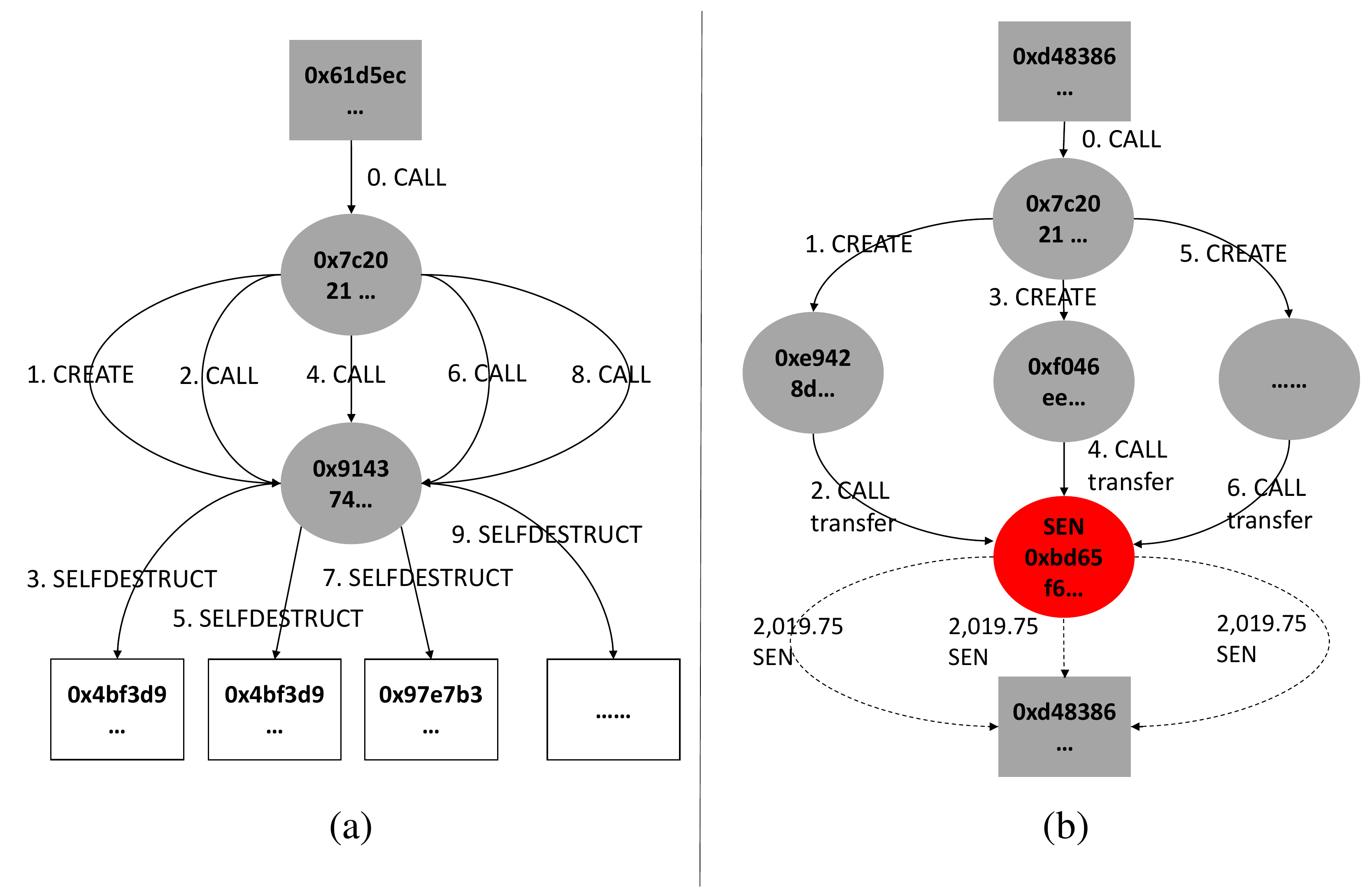}
    \caption{The dynamic call graph of a \textit{suicide bomb} DoS attack and an ERC20 airdrop hunting attack. Square: EOA, Circle: smart contract; Grey Box: attacker, Red Box: victim;
    Solid Line: transaction, Dotted Lines: ERC20 token transfer; The number before the opcode is the execution order for each opcode.}
    \label{fig:eva_dos&airdrop_callgraph}
\end{figure}

	\begin{itemize}[leftmargin=*]
		\item \textit{Suicide bomb DoS attack.}\tab
		From blocks ranging from $2,000,000$ to $3,000,000$, there exists a smart contract~\footnote{\scriptsize{0x7c20218efc2e07c8fe2532ff860d4a5d8287cb31}} that contributes $34,148$ and 
		$33,980$ times of smart contract creation and self-destruction,   respectively. 
		The only functionality of the newly created smart contract is to self-destruct itself, and transfer its balance ($1$ Wei or $0$ Wei) to a non-existent account. 
		
		We take a transaction~\footnote{\scriptsize{0xa02be5a3f2687b68e4643e73d26c4661dc66fb3550aa34fc9-
				6abfa4bcb0bf8b6}} as an example to illustrate its purpose. 
		Fig.~\ref{fig:eva_dos&airdrop_callgraph}(a) shows the dynamic call graph. The EOA (\code{0x61d5ec}) first invokes (index $0$) a smart contract (\code{0x7c2021}) to create (index $1$)
		a very simple contract (\code{0x914374}). Its functionality is to self-destruct itself, and transfer its balance ($0$ Wei in this example) to a non-existent account. 
		
		For simplicity, we only draw the first ten transactions, and this normal transaction actually triggered $320$ times of self-destruction.
		
		It is worth noting that, the destruction of a smart contract actually happens only when the execution of the normal transaction that initiates these internal transactions finishes (index $0$). Thus, the contract \code{0x914374} can execute the opcode \code{SELFDESTRUCT} multiple times before it is actually self-destructed. Moreover, according to the definition of the opcode \code{SELFDESTRUCT}, it will create a new EOA account (Section~\ref{subsec:b_accounts}), without paying for the $25,000$ gas charge~\footnote{This vulnerability has been fixed in the EIP150~\cite{eip-150} hard fork of Ethereum}, which is the gas needed to create a new account. These newly created accounts will consume lots of storage resources on the blockchain. This is called the \textit{suicide bomb} DoS attack~\cite{chen2017DoS}.
		
		\smallskip
		\item \textit{Airdrop hunting attack.}\tab
		In blocks ranging from $6,500,000$ to $7,500,000$, there is a smart contract account~\footnote{\scriptsize{0xe9428d4a341ac20e9f2e6b95b12c9ad52733fcd9}} that contributes $501,919$ creation and $526,079$ times of smart contract creation and self-destruction, respectively. 
		We randomly pick a normal transaction~\footnote{\scriptsize{0x5a5fb2f3d097c44d0454612404097eb51f0025bf86c5f25e1902639e139b944b}}, and draw the 
		dynamic call graph in Fig.~\ref{fig:eva_dos&airdrop_callgraph}(b) to help us understand its purpose. 
		As shown in the graph, the smart contract (\code{0x7c2021}) continually creates new smart contracts to transfer $2,019.75$ SEN tokens to the EOA (\code{0xd48386}) that initiates this transaction. 
		The SEN token has an aggressive marking strategy, which will reward a few tokens for every \textit{new} account that has made a 
		transaction with SEN. This strategy is adopted by many token smart contracts.
		The purpose of creating so many \textit{new} smart contracts is abusing this strategy to obtain rewards. Destroying these new smart contracts is 
		not necessary but can save cost. 
		This kind of rewards is usually called airdrop reward. Therefore, this attack is called \textit{airdrop hunting} attack.
		
		\begin{figure}[t]
	\centering
	\includegraphics[width=0.5\textwidth]{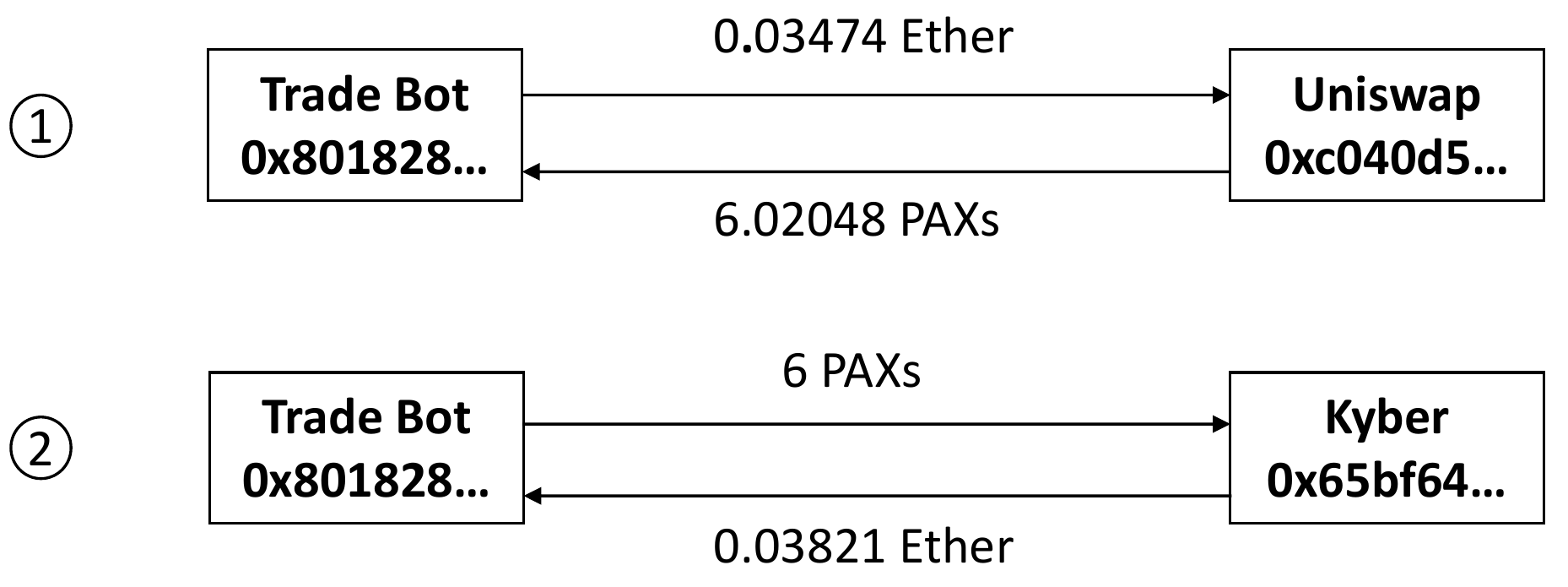}
	\caption{The trades in an arbitrage (normal transaction). The number in circle represents the execution order.}
	\label{fig:eva_arbitrage_assets_flow}
\end{figure}
		\smallskip
		\item \textbf{Automated arbitrage trading. }
		In blocks ranging from $8,900,000$ to $10,110,000$, there is a
		smart contract account~\footnote{\scriptsize{0x8018280076d7fa2caa1147e441352e8a89e1ddbe}} that contributes $510,390$ creation and $537,992$ times of smart contract creation and self-destruction, respectively.
		After analyzing the suspicious transactions, we find this is a \textit{trade bot}, which buys and sells digital assets among decentralized exchanges using arbitrage. Though this cannot be considered as an attack, this still shows the capability of our system to understand the behaviors of smart contracts.

		Fig.~\ref{fig:eva_arbitrage_assets_flow} shows the digital assets transfer in an 
		arbitrage (normal) transaction~\footnote{\scriptsize{0x3cf41ad4f703fe61368139b8482e75de53a335b9d76039ca071530bb5292b0c7}}, which includes two trades. The \textit{trade bot} (\code{0x801828}) first exchanges $6.02048$ PAXs~\cite{PAX} with 
		$0.03474$ Ether from Uniswap~\cite{Uniswap}, and then exchanges $0.03821$ Ether with $6$ PAXs from Kyber~\cite{kyber}. 
		As a result, the \textit{trade bot} gets $0.003$ Ether and $0.02$ PAXs due to the exchange rate differences between the two exchanges Kyber and Uniswap.

		We further analyze the purpose of the self-destruction of smart contracts. The \textit{trade bot} first created lots of smart contracts in advance with lower gas price. When performing arbitrage, attackers will set up a higher gas price so that their trade transactions have a higher priority when being packed. That's because miners tend to pack transactions with higher gas price. After that, they self-destruct the smart contracts to receive the returned gas at a higher gas price since the current gas price used in the transaction is high.

	\end{itemize}
	
	\subsection{Database Indices}
	
	Table~\ref{tab:app_elasticindices} shows the database indices used in \aggregator{}. It is similar with the schema of the relational database.

	\begin{table*}
    \centering
    \caption{ElasticSearch Indices}
    \label{tab:app_elasticindices}
    \footnotesize
    \begin{threeparttable}
     \resizebox{.75\textwidth}{!}{
    \begin{tabular}{|l|l|l|l|l|}
    \hline
    \textbf{Index Name} & {Field} & \minitab{Field of \\ Nested Field} & \minitab{Field of \\ Nested Field of \\ Nested Field} & \minitab{Field of \\ Nested Field of \\ Nested Field of \\ Nested Field}\\
    \hline
    \multirow{10}{1in}{Block} & Difficulty$^R$ & & & \\
    & ExtraData & & & \\
    & GasLimit$^R$ & & & \\
    & GasUsed & & & \\
    & Hash$^R$ & & & \\
    & Miner$^R$ & & & \\
    & Number$^R$ & & & \\
    & Timestamp$^R$ & & & \\
    & TxnCount\bigstrut & & & \\\cline{2-3}
    & \multirow{16}{1in}{Transaction} & CallFunction$^R$ & & \\
    & & ConAddress & & \\
    & & CumGasUsed & & \\
    & & FromAddress$^R$ & & \\
    & & GasLimit$^R$ & & \\
    & & GasPrice$^R$ & & \\
    & & GasUsed$^R$ & & \\
    & & GetCodeList$^R$ & & \\
    & & Hash$^R$ & & \\
    & & IntTxnCount & & \\
    & & Nonce$^R$ & & \\
    & & Status & & \\
    & & ToAddress$^R$ & & \\
    & & TxnIndex & & \\
    & & Value$^R$\bigstrut & & \\\cline{3-4}
    & & \multirow{11}{1in}{InternalTxns} & CallFunction & \\
    & & & CallParameter & \\
    & & & ConAddress & \\
    & & & EvmDepth & \\
    & & & FromAddress & \\
    & & & GasLimit & \\
    & & & Output & \\
    & & & ToAddress & \\
    & & & TxnIndex & \\
    & & & Type & \\
    & & & Value\bigstrut & \\\cline{3-4} 
    & & \multirow{3}{1in}{Logs} & Address & \\
    & & & Topics & \\
    & & & Data\bigstrut & \\\cline{3-4}
    & & \multirow{6}{1in}{ReadCommittedState$^R$} & Address & \\
    & & & Balance & \\
    & & & CodeHash & \\
    & & & CodeSize & \\
    & & & Nonce\bigstrut & \\\cline{4-5}
    & & & \multirow{2}{1in}{Storage} & Key \\
    & & & & Value\bigstrut \\\cline{3-5}
    & & \multirow{5}{1in}{ChangedState} & Address & \\
    & & & Balance & \\
    & & & Nonce\bigstrut & \\\cline{4-5}
    & & & \multirow{2}{1in}{Storage} & Key \\
    & & & & Value\bigstrut \\\cline{3-5}
    \hline
    \multirow{8}{1in}{Code} & Number & & & \\
    & Timestamp\bigstrut & & & \\\cline{2-3}
    & \multirow{4}{1in}{Transaction} & Hash & & \\
    & & TxnIndex & &\\
    & & Input$^R$\bigstrut & & \\\cline{3-4}
    & & \multirow{3}{1in}{Contract$^R$} & Address & \\
    & & & Hash & \\
    & & & Code\bigstrut & \\
    \hline
    \multirow{7}{1in}{State} & Number & & & \\
    & Timestamp\bigstrut & & & \\\cline{2-3}
    & \multirow{5}{1in}{Transaction} & Hash & & \\
    & & TxnIndex & & \\
    & & Create & & \\
    & & Reset & & \\
    & & Suicide$^R$\bigstrut & & \\
    \hline
    \end{tabular}}
    \begin{tablenotes}
        \footnotesize
        \item[$^R$]: fields that are necessary for replaying transactions.
    \end{tablenotes}
    \end{threeparttable}
\end{table*}

\end{document}